\documentclass[aps,prx,twocolumn,superscriptaddress,showpacs]{revtex4-2}
\usepackage{hyperref}
\usepackage[dvips]{graphicx}
\usepackage{amssymb,amsmath,amsfonts}
\usepackage[T1]{fontenc}
\usepackage{float}
\usepackage{color,soul}
\usepackage{booktabs}
\usepackage{changes}
\usepackage{braket}
\usepackage{physics}
\usepackage{tabularx,multirow}
\usetikzlibrary{automata}
\hypersetup{
    colorlinks,
    linkcolor={red!80!black},
    citecolor={blue!80!black},
    urlcolor={blue!80!black}
}

\usepackage{graphicx}

\usepackage{enumerate}
\usepackage{comment}
\usepackage{tikz}
\usetikzlibrary{decorations.pathreplacing}

\newcommand{\cq}{\mathcal{Q}}
\newcommand{\cj}{\mathcal{J}}

\begin{abstract}

We characterize steady-state static and dynamic properties in a broad class of mass transport processes on a periodic hypercubic lattice of volume $L^d$, where both mass and {\it center-of-mass} (CoM) remain conserved and detailed balance is violated in the bulk; we specifically consider the models in $d=1$ and $2$ dimensions. Using a microscopic approach, we exactly determine the decay (or, growth) exponents for various dynamic and static correlation functions. 
We show that, despite constrained dynamics due to the CoM conservation (CoMC), the density relaxation is indeed diffusive. However, fluctuation properties are strikingly different from that in diffusive systems with a {\it single} (mass) conservation law as dynamic and static fluctuations are more suppressed in systems with CoM conservation, resulting in an extreme (``class-I'') hyperuniformity in certain cases. 
In the thermodynamic limit, the steady-state variance $\langle {\cal Q}^2(T) \rangle_c$ of time-integrated bond current ${\cal Q}(T)$ across a bond in time interval $T$ exhibits the following long-time behavior: $\langle {\cal Q}^2(T) \rangle_c \simeq A_1 T + A_2 + A_3 T^{-d/2}$. 
The exponents governing the small-frequency behavior of the power spectra $S_J(f) \simeq A_1+ {\rm Const.} f^{\psi_J}$ for bond current are exactly determined as $\psi_J=3/2$ and $2$ in $d=1$ and $2$ dimensions, respectively; the corresponding unequal-time current-current correlation functions decay as $t^{-5/2}$ and $t^{-3}$ as a function of time $t$ in $d=1$ and $2$ dimensions, respectively.
Remarkably, depending on dimensions and microscopic details, the prefactor $A_1$ can vanish, causing the variance to eventually {\it saturate}, implying a class-I ``dynamic hyperuniformity''.
We also compute the static structure factor $S(q)$, which varies as the square of wave number $q$ in the small-$q$ limit, i.e., $S(q) \sim q^2$, implying a class-I spatial hyperuniformity. 
\end{abstract}

\begin{document}

\title{Hyperuniformity in mass transport processes with center-of-mass conservation: Some exact results}

\author{Animesh Hazra}
\email{animesh\_edu@bose.res.in}
\affiliation{Department of Physics of Complex Systems, S. N. Bose
   National Centre for Basic Sciences, Block-JD, Sector-III, Salt Lake, Kolkata 700106, India.}
\author{Anirban Mukherjee}
\affiliation{Department of Physics of Complex Systems, S. N. Bose
   National Centre for Basic Sciences, Block-JD, Sector-III, Salt Lake, Kolkata 700106, India.}
   \affiliation{Institute of Physics, Academia Sinica, Taipei 11529, Taiwan.}
\author{Punyabrata Pradhan}
\affiliation{Department of Physics of Complex Systems, S. N. Bose
   National Centre for Basic Sciences, Block-JD, Sector-III, Salt Lake, Kolkata 700106, India.}

\date{\today}
\maketitle

\section{Introduction}

\label{Introduction}

Recently, there has been a surge of interest in characterizing a remarkable state of matter that is amorphous or disordered, yet correlated and known as ``hyperuniform'' matter  \cite{Torquato2018Jun}. While they lack true long-range order, these many-body systems exhibit long-wave-length density fluctuations, that are  anomalously suppressed. First, consider an equilibrium disordered fluid with density $\rho$ in $d$-dimensional space.
In that case, the variance of particle number, or mass, $M_l$ in a large subsystem of volume $l^d$ is equal to $\sigma^2(\rho) l^a$, with $a=d$ following the central limit theorem, where $\sigma^2(\rho)$ is the scaled variance of subsystem mass and is proportional to the compressibility. 
By tuning parameters like temperature and pressure, one can have a state with the exponent $a$ being larger than the dimension ($a>d$); an example of this scenario is an equilibrium critical point (say, that of water and vapor), where fluctuation, or, equivalently, the compressibility, diverges in the thermodynamic limit. Indeed, the opposite also can happen where, in the presence of interactions, particles in a system can dynamically organize themselves in such a fashion that the subsystem-mass fluctuations are greatly reduced as the subsystem size increases, leading to the vanishing compressibility; these systems are ``hyperuniform'' as $(d-1) \le a < d$ \cite{Torquato2003Oct}. In fact, a perfectly {\it ordered} crystalline state (achieved only at zero temperature) is hyperuniform where $a=d-1$; 
other examples include one-component Coulomb gas \cite{Baus1978Dec, Tamashiro1999Jun} and jammed granular matter \cite{Torquato2003Oct, Donev2005Aug}, which are incompressible and hyperuniform.
Notably, no {\it disordered} systems, with short-ranged interactions and at finite temperature, are however known to exhibit hyperuniformity in equilibrium;
rather, for such systems, hyperuniformity has so far been observed only in nonequilibrium settings \cite{Corte2008May, Berthier2011Mar, Zachary2011Apr, Wilken2020Sep, Wilken2023Aug, Salvalaglio2020Sep, Maire2024Aug, Lei2019Nov, Lei2019Jan}.
As discussed above, spatial correlations within a system are quantified by space-integrated (extensive) observables such as subsystem particle number. Similarly, time-integrated observables, such as cumulative particle flux ${\cal Q}(T)$ across a surface in the time interval $T$, can be used to characterize dynamic correlations.
Typically, in one-dimensional diffusive systems with a {\it single} (say, mass) conservation law, the variance ${\cal Q}^2(T)$  of time-integrated current in the thermodynamic limit grows sublinearly with time $T$, i.e., the temporal growth of ${\cal Q}^2(T) \sim T^{b}$ is subdiffusive with $b=1/2$ \cite{DeMasi2002May, Derrida2009Jul, Mukherjee2023Feb, Hazra2024Aug}; near criticality though, the exponent $b$ can take other values, less \cite{Mukherjee2023Feb, Mukherjee2024Aug} or greater (but, still $b<1$) \cite{Chakraborty2024Nov} than one-half. Similar anomalous growth of time-integrated fluctuations can also happen for various other quantities, such as cumulative displacement of a tagged particle in single-file diffusion where $b=1/2$ \cite{Barkai2009Feb} or cumulative activity (or, the avalanche size) in one-dimensional sandpiles where, quite interestingly, $b=0$ \cite{Garcia-Millan2018Jul}; the latter case ($b=0$) in fact provides an example of an extreme form of hyperuniformity in time domain, analogous to hyperuniformity exponent $a=d-1$ in space domain. Indeed, the subdiffusive ($b<1$) temporal growth of the variance of a time-integrated quantity is a manifestation of long-ranged temporal correlations in the system and is called ``dynamic hyperuniformity'', the notion recently introduced in the context of threshold-activated systems like sandpiles \cite{Garcia-Millan2018Jul, Mukherjee2023Feb}. Exact characterization of such correlations in interacting-particle systems is of relevance also from a general theoretical point of view and has been done in this paper for current fluctuations in nonequilibrium mass transport processes with center-of-mass (CoM) conservation. 



Two celebrated models for disordered nonequilibrium systems are the conserved (``fixed-energy'') Manna sandpiles \cite{Manna1991Apr, Dickman2001Oct} and the biased random organization (BRO) \cite{Corte2008May, Wilken2021Jul}. Both the models have been studied in a variety of contexts in the past, including absorbing-phase transition (APT) \cite{Vespignani1998Dec, Dickman2001Oct}, reversible-irreversible transition in sheared colloids \cite{Wilken2020Sep, Ma2019Feb}, random close packing (RCP) in disordered solids and jamming phenomenon \cite{Corte2008May, Corte2009Dec, Menon2009Jun, Torquato2000Mar, Donev2005Aug, Tjhung2015Apr, Maher2023Dec}, etc. 
Although hyperuniform structures have been envisaged since a long time  \cite{Torquato2003Oct}, it was only recently observed that these two models exhibit hyperuniform fluctuations in their (quasi-)nonequilibrium steady states \cite{Hexner2015Mar, Hexner2017Jan, Grassberger2016Oct, Wiese2024Aug, Dandekar2020Dec}. 
In the BRO model, provided that they overlap, a pair of particles of, say, unit diameter are given a random kick (displacement) of {\it identical} magnitude, but in the diametrically {\it opposite} directions, ensuring that their center of mass remains the same. Consequently, there are two conserved quantities in the system: Total mass and center-of-mass (CoM). Moreover, the microscopic dynamics in the bulk break time-reversal symmetry and the BRO model exhibits hyperuniformity in the so-called ``active'' phase, which is analogous to the ``irreversible'' phase of the colloidal system investigated in the experiment \cite{Wilken2021Jul, Ma2019Feb}. 
Perhaps not surprisingly, the BRO model is extremely challenging to solve analytically, even in one dimension, due to the nontrivial overlapping condition, which allows for particles hopping only when the particles overlap with each other. 
Indeed, lately there is a growing interest to develop a rigorous theoretical understanding of how multiple conservation laws determine large-scale properties, concerning density relaxation and current and mass fluctuations, of minimal model systems \cite{Spohn2012Dec, Spohn2020Sep}.  In such a scenario, it is highly desirable to examine CoM-conserving models with simpler dynamical rules, allowing the systems to be rigorously dealt with through a microscopic approach.

In one dimension, if the overlapping condition is relaxed and the random displacements depending on the inter-particle gaps are chosen appropriately, the modified model is related to another class of widely studied many-body systems, known as random average processes (RAPs) \cite{Liggett, Ferrari1998Jan, Krug2000Apr,Rajesh2000May}, which consist of interacting particles diffusing on a circle (continuum), but now obeying the (additional) CoM conservation (CoMC). While the particles remain correlated, resulting in a nontrivial spatial structure in the systems, these one-dimensional models allow explicit analytical calculations. 
For simplicity, in the rest of the paper, we focus on the unbounded versions of these mass transport processes, known as {\it mass chipping models} (MCMs) \cite{Liu1995Jul, Bondyopadhyay2012Jul, Chakraborti2000Sep, Patriarca2005, Yakovenko2009Dec, Das2016Jun, Carinci2013Aug, Redig2017Oct, vanGinkel2016Apr}, which are
the generalized variants of the Kipnis-Marchioro-Presutti (KMP) models  \cite{Kipnis1982Jan,  Aldous1995Jun} and, in one dimension, can be obtained through an exact mapping  \cite{Krug2000Apr, Evans2005Apr} of a RAP of $L$ particles on a circle to a system  on a ring of $L$ discrete lattice sites having continuous mass. 
That is, a particle in the RAP can be thought of as a lattice site, and the inter-particle gap is a (continuous) non-negative mass variable assigned to the site; also the dynamical mass transfer rules in the RAP have one-to-one correspondence to the mapped mass transport process.  
In a related context, an unbounded version of a RAP with an overlapping (threshold) condition, which allows movement of two particles on a line only when the gap between the particles exceed a certain threshold value (the condition differs slightly from that in the BRO model), has received a lot of attention in the last decade. The one-dimensional model is a continuous-mass variant of the discrete-mass Manna sandpile \cite{Manna1991Apr, Dickman2001Oct}, which has been intensively studied to characterize APT with a conserved (density) field \cite{Basu2012Jul, Mukherjee2023Feb}. In the MCMs explored here, we do not however impose any threshold condition in microscopic dynamics to make the systems analytically tractable, and consequently the phase transition is lost.

The MCMs break time-reversal symmetry (violating detailed balance) in the bulk for generic parameter values and, consequently, the steady-state probabilities of the microscopic configurations in most cases are not described by the equilibrium Boltzmann-Gibbs distribution and not a-priori known. 
Throughout the past several decades, they have been extensively studied in the literature in various contexts, such as cloud formation \cite{Friedlander1977Sep}, force fluctuation in jammed granular media \cite{Liu1995Jul}, wealth distribution in a population \cite{Chakraborti2000Sep, Patriarca2005,Chatterjee2004Apr}, and traffic flow \cite{Chowdhury2000May}, etc. 
However, the issue concerning the precise role of an additional conserved quantity, such as CoM, on the relaxation and (static and dynamic) fluctuation properties of these systems has yet to be explored. Indeed, the problem is of general relevance, as there has recently been a surge of interest in characterizing the large-scale properties of systems with multiple conservation laws \cite{Spohn2022Mar, Spohn2020Sep, Boldrighini1983Jun,Castro-Alvaredo2016Dec,Ogunnaike2023Nov,Feldmeier2020Dec}, notably in the context of quantum many-body systems like fractonic fluids \cite{Morningstar2020Jun, Shenoy2020Feb, Gromov2020Jul}.

In this paper, using a microscopic approach, we analytically calculate various static and dynamic correlation functions in a broad class of MCMs having both mass and CoM conservation in $d=1$ and $2$ dimensions (the results can be suitably generalized to arbitrary dimensions). We find that the density relaxation, despite the dynamics being strongly constrained due to the CoM conservation (CoMC), is in fact diffusive. To substantiate this assertion, we consider, for simplicity, an infinite domain and consider density relaxation by taking a step-like initial density profile $\rho(X,t=0) = \rho_{in}(X) =\rho_0+\rho_1$ for $x<0$ and $\rho_0$ otherwise. We show the time-dependent relative density profile $\rho(X,t)-\rho_0 = {\cal R}(X/\sqrt{t})$ obeys a diffusive scaling, where ${\cal R}(z)$ is a scaling function, with the scaled argument $z=X/\sqrt{t}$, and is determined exactly.  However, fluctuation properties are strikingly different from that in diffusive systems with a single (mass) conservation law.
In the thermodynamic limit, the steady-state  second cumulant, or the variance, $\langle {\cal Q}^2(T) \rangle_c = \langle {\cal Q}^2(T) \rangle - \langle {\cal Q}(T) \rangle^2$ of the time-integrated bond current ${\cal Q}(T)$ across a bond in the time interval $T$ has the following long-time behavior: $\langle {\cal Q}^2(T) \rangle_c = \langle {\cal Q}^2(T) \rangle \simeq A_1 T + A_2 + A_3 T^{-d/2}$ (note that $\langle {\cal Q}^2(T) \rangle_c = \langle {\cal Q}^2(T) \rangle$ as the average bond current $\langle {\cal Q}(T) \rangle = 0$ is zero in steady state). 
The exponents governing the small-frequency behavior of the power spectra $S_J(f) \simeq A_1+ {\rm Const.} f^{\psi_J}$ for bond current are exactly determined as $\psi_J=3/2$ and $2$ in $d=1$ and $2$ dimensions, respectively; in $d$ dimensions, we have $\psi_J=(1+d/2)$.
Remarkably, depending on dimensions and microscopic details, the coefficient $A_1$ can vanish (e.g., for $d=1$). 
That is, the variance of time-integrated bond current in the time interval $T$ eventually {\it saturates} as a function of time $T$ to a density-dependent coefficient $A_2(\rho)$, implying a ``dynamic hyperuniformity''; in that case, we have $\langle {\cal Q}^2(T) \rangle_c \sim T^{b}$ with $b=0$, which signifies an extreme form of hyperuniformity in temporal domain observed previously for avalanche time-series in a one-dimensional sandpile \cite{Garcia-Millan2018Jul}.  
Furthermore, the long-time (or, equivalently, small-frequency) asymptotics for dynamic (two-point and unequal-time) ``cross"-correlation  functions for the bond currents in orthogonal directions (e.g., correlations between currents along $x$ and $y$ directions) decay as a power law, where the exponents of the power laws are exactly determined.
We also calculate static structure factor $S(q)$, which varies as the square of wave number $q$ in the small-$q$ limit, i.e., $S(q) \sim q^2$ as $q \to 0$.
This is a direct consequence of the fact that, in the thermodynamic limit, the spatial integral of the two-point density correlation function for mass is zero in these systems.

We organize the paper as follows. In Sec. \ref{Sec:MCM_CoM_I_1D}, we introduce a CoM-conserving mass chipping model (MCM), called MCM-CoMC, in one dimension and demonstrate its equivalence to a CoM-conserving random average process (RAP). In Sec. \ref{Sec:Hydro_MCM_COMI1D}, we derive hydrodynamics of the one-dimensional MCM-CoMC with nearest-neighbor mass transfer; in Sec. \ref{Sec:1D_MCM_COM_I}, we explore dynamic correlations for the model. In Sec. \ref{Sec:1D_MCM_COM_II}, we explore a one-dimensional variant of the MCM-CoMC, which now have microscopic dynamics with finite-range mass transfer. In Secs. \ref{Sec:2D_MCM_COM_I} and \ref{Sec:2D_MCM_COM_II}, we introduce a couple of two-dimensional versions of the MCM-CoMCs. We study the asymptotic behavior of the MCM-CoMCs in higher dimensions in Sec. \ref{Sec:dD_MCM_COM}. Finally, in Sec. \ref{Sec:Conclusion}, we summarize the paper with some concluding remarks.

\section{One-dimensional MCM-CoMC I: Nearest-neighbor mass transfer}\label{Sec:MCM_CoM_I_1D}

\begin{figure}[ht!]
\centering
    \begin{tikzpicture}
    \def\a{4} 
    \def\b{2.0} 
    \def\radius{0.2} 

    \draw[line width =0.8mm] (0,0) ellipse (\a cm  and \b cm);
    \pgfmathsetseed{1} 
    \foreach \i in {0,1,...,5} {
        \pgfmathsetmacro\angle{360/6 * \i}
        \pgfmathsetmacro\x{\a * cos(\angle)}
        \pgfmathsetmacro\y{\b * sin(\angle)}
        \filldraw (\x,\y ) circle (\radius);
        \pgfmathsetmacro\thenum{1.2 + rand*(0.6 - 0.4)}
        \draw[fill=purple] (\x-0.3,\y) rectangle (\x+0.3,\y+\thenum);
    }
    \draw[fill=green] (1.7,3.2) rectangle (2.3,3.7);
    \draw[fill=blue] (1.7,3.9) rectangle (2.3,4.4);
    \draw[->,line width=0.6mm, purple] (2.4,4.2) -- (3.4,4.2);
    \draw[->,line width=0.6mm, purple] (1.6,3.45) -- (0.6,3.45);
    \draw [ultra thick, decorate,decoration={brace,amplitude=6pt,mirror}] (2.4,1.65) -- (2.4,2.95);
    \node at (3.2,2.4) {\textbf{$m_i\xi_i$}} ;
    \node at (2.8,4.6) {\textbf{$\frac{m_i\tilde{\xi}_i}{2}$}} ;
    \node at (1.1,3.9) {\textbf{$\frac{m_i\tilde{\xi}_i}{2}$}} ;
    \node at (2,1.3) {$i$};
    \node at (3.4,-0.4) {$i-1$};
    \node at (-2.0,1.3) {$i+1$};
\draw[line width =0.8mm] (0,8) ellipse (4 cm  and 2.0 cm);
    \foreach \i in {0,1,...,6} {
        \pgfmathsetmacro\angle{360/6 * \i}
        \pgfmathsetmacro\x{\a * cos(\angle)}
        \pgfmathsetmacro\y{\b * sin(\angle)}
        \filldraw (\x,\y+8 ) circle (0.3);
    }
    \node at (0.0,10.9) {$m_i$};
    \draw [ultra thick, purple, decorate,decoration={brace,amplitude=10pt}] (-2,10.3) -- (2,10.3);
    \begin{scope}[rotate=10]
        \draw [ultra thick, green, decorate,decoration={brace,amplitude=4pt,mirror}] (-0.2,9.6) -- (1.0,9.6);
    \end{scope}
    \begin{scope}[rotate=-10]
        \draw [ultra thick, blue, decorate,decoration={brace,amplitude=4pt,mirror}] (-1.0,9.6) -- (0.3,9.6);
    \end{scope}
    \node at (-1.2,9.0) {\textbf{$\frac{m_i\tilde{\xi}_i}{2}$}} ;
    \node at (1.2,9.0) {\textbf{$\frac{m_i\tilde{\xi}_i}{2}$}} ;
    \draw[ultra thick, dashed, green] (-0.8,10 ) circle (0.3);
    \draw[ultra thick, dashed, blue] (0.8,10 ) circle (0.3);
    \begin{scope}[rotate=10]
        \draw[->,line width=0.6mm, red] (0.0,10.2) -- (0.8,10.2);
    \end{scope}
    \begin{scope}[rotate=-10]
        \draw[->,line width=0.6mm, red] (0.0,10.2) -- (-0.8,10.2);
    \end{scope}
    \draw[<->,line width=1.9mm, violet] (-4.0,2.0) -- (-4.0,6.0);
    \node[rotate=90] at (-3.5, 4) {\textbf{Equivalence}};
    \node at (0, 8) {\textbf{RAP with CoM conservation}};
    \node at (0.0, 0.2) {\textbf{MCM with CoM conservation}};
    \node at (3.4, 7.6) {$X_{i-1}$};
    \node at (2.7, 10) {$X_{i}$};
    \node at (-2.8, 10) {$X_{i+1}$};
\end{tikzpicture}
\caption{\textit{Equivalance between random average process (RAP) and mass chipping model (MCM) with CoM conservation (CoMC) in one dimension through a schematic diagram.} Top panel -- \textit{RAP with CoMC}: Two particles (black circles) located at $X_i$ and $X_{i+1}$ are displaced in the opposite directions (inward) by an equal amount, i.e., $m_i \tilde{\xi}_i/2$, where $m_i \equiv X_{i+1} - X_i$ is the inter-particle distance, or gap, between $i$th and $(i+1)$th particles. Here $\tilde \xi_i=(1-\xi_i)$ and $\xi_i$ is a random fraction uniformly distributed in  $[0,1]$. Bottom panel -- \textit{MCM with CoMC}: A site (represented by a black circle) contains a certain amount of (continuous) mass (dark-violet rectangular bars). During a mass transfer process, a site retains a random fraction $\xi_i$ of its mass $m_i$, and the remaining fraction of mass, $(1 - \xi_i)m_i$, is \textit{equally} divided into two parts, with one being transferred to its left neighbor (green rectangle) and the other to its right neighbor (blue rectangle). In that case, cumulative (time-integrated) current increases by $m_i \tilde \xi_i/2$ across bond $(i,i+1)$ and decreases by the same amount across the other bond $(i-1,i)$. } 
    \label{fig:model}
\end{figure}
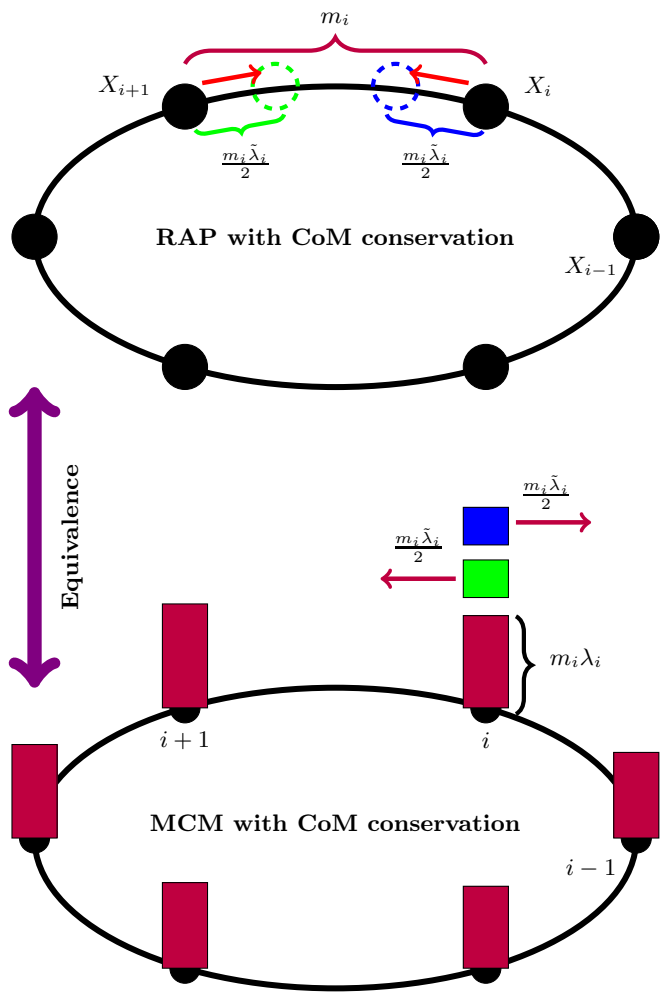

In this section, we explore a broad class of conserved-mass transport processes, called mass chipping models (MCMs) \cite{Chakraborti2000Sep, Yakovenko2009Dec, Bondyopadhyay2012Jul, Das2015Nov, Hazra2024Aug}, where masses fragment, diffuse, and coalesce with the neighboring masses at a constant rate. 
Because of their analytical tractability, these processes, which are generalized variants of the well-known random average processes (RAPs) \cite{Liggett, Ferrari1998Jan, Krug2000Apr,Rajesh2000May} or the Kipnis-Marchioro-Presutti (KMP) models \cite{Kipnis1982Jan,  Aldous1995Jun}, have been studied mainly in one dimension in the past. 
Notably, on a periodic domain, unlike other lattice gases, such as symmetric simple exclusion processes (SSEP) \cite{Derrida2009Jul, Grabsch2023Apr} and zero-range processes (ZRPs) \cite{Evans2005Apr,Harris2005Aug}, MCMs generally violate detailed balance in the bulk due to the broken time-reversal symmetry; consequently, their steady-state measures are in most cases not described by the equilibrium Boltzmann-Gibbs distribution. 
The detailed definition of a broad class of MCMs, and their connection to other models such as RAPs, can be found in Ref. \cite{Hazra2024Aug}, where relaxation and fluctuation properties have been studied.
However, in this work, we impose an {\it additional} conservation law on the microscopic dynamics, where both mass and center-of-mass (CoM) now remain conserved.

We introduce a model of {\it MCM with CoM conservation} (MCM-CoMC) and study the role of CoM conservation on the time-dependent (density relaxation and dynamic fluctuation) properties of the system.
Let us begin by considering a system on a one-dimensional periodic lattice of size $L$, where lattice sites are labeled as $i = 0, 1, \ldots, L-1$. A site $i$ is associated with a continuous mass variable $m_i \geq 0$, with the total mass $M=\sum_{i=0}^{L-1} m_i = \rho L$ fixed and $\rho$ being the global density; defining similar CoM-conserving models in higher dimensions $d$ is straightforward and the MCM-CoMCs in $d=2$ are studied later. We consider a continuous-time process, where mass $m_i$ at site $i$ gets chipped off with unit rate. A random fraction $\xi_i$ of mass $m_i$ is retained and the remaining fraction, $(1-\xi_i)$, of mass, then gets fragmented in two {\it equal} halves; one half $(1-\xi_i) m_i/2$ coalesce with the right-neighbor ($i+1$) mass and the other half coalesce with the left-neighbor ($i-1$)  mass. Evidently, the mass transfer rule locally conserves {\it mass} as well as \textit{center of mass} (CoM). In Fig. \ref{fig:model}, we demonstrate the equivalence between two models: the random average process (RAP) with CoM conservation and the mass chipping model (MCM) with CoM conservation. In top panel, we describe a \textit{RAP with CoMC}. In this case, two particles (black circles) located at $X_i$ and $X_{i+1}$ are displaced in the opposite directions (inward) by an equal amount, i.e., $m_i \tilde{\xi}_i/2$, where $m_i \equiv X_{i+1} - X_i$ is the inter-particle distance, or gap, between $i$th and $(i+1)$th particles. In bottom panel, we describe a \textit{MCM with CoMC}. In this case, a site (represented by a black circle) contains a certain amount of (continuous) mass (dark-violet rectangular bars). During a mass transfer process, a site retains a random fraction $\xi_i$ of its mass $m_i$, and the remaining fraction of mass, $(1 - \xi_i)m_i$, is \textit{equally} divided into two parts, with one being transferred to its left neighbor (green rectangle) and the other to its right neighbor (blue rectangle).

For the MCM-CoMC in one dimension, the time evolution of mass at a site $i$ can be expressed in terms of the following infinitesimal-time update rules:
\begin{align}\label{eq:mass_update}
m_i(t+dt)=
    \begin{cases}
        \textbf{event}         & \textbf{prob.} \\
        m_i(t)-\tilde{\xi}_i m_i(t)       & dt\\
        m_i(t)+\frac{\tilde{\xi}_{i+1}}{2} m_{i+1}(t)   & dt\\
        m_i(t)  +\frac{\tilde{\xi}_{i-1}}{2} m_{i-1}(t)   & dt\\
        m_i(t)  & (1-3dt),
        \end{cases}
\end{align}
where $\tilde{\xi_i}=1-\xi_i\in (0,1)$ is a random variable, drawn from a distribution $\phi(\xi_i)$, with the first and the second moments being denoted as $\langle \xi_i\rangle =\mu_1$ and $\langle \xi_i^2\rangle= \mu_2$, respectively. In this paper, we consider, for simplicity, a uniform distribution $\phi(\xi_i) = 1$, with $\mu_1 = 1/2$ and $\mu_2 = 1/3$.

\subsection{Hydrodynamics}
\label{Sec:Hydro_MCM_COMI1D}

In this section, we investigate the temporal evolution of a density perturbation and its relaxation dynamics.
According to the update rule Eq. (\ref{eq:mass_update}), the time  evolution of local mass at site $X$ can be written as
\begin{align}
\label{eq:mass_upd}
    \frac{d\langle m_X(t) \rangle}{dt} = D\left(\langle m_{X+1}(t)-2m_X(t)+m_{X-1}(t) \rangle\right).
\end{align}
Here, $D=\mu_1/2$ is the bulk-diffusion coefficient, which is {\it independent} of density. Provided we consider density relaxation on a ring of $L$ sites, we can scale position $X$ and time $t$ as $x=X/L$ and  $\tau=t/L^2$, and we obtain, from Eq. \eqref{eq:mass_upd}, 
\begin{align}
    \frac{\partial h(x,\tau)}{\partial\tau} = D \frac{\partial^2 h(x,\tau)}{\partial x^2},
\end{align}
which governs the time-evolution equation for scaled density field $h(x=X/L,\tau=t/L^2) =\langle m_X(t) \rangle \equiv \rho_X(t)$.
The aforementioned equation is a linear partial differential equation, which can be solved for a given initial condition. 
For simplicity, to show the diffusive scaling property of time-dependent density profile, we take an initial density profile being a step-like profile $\rho_X(t=0) = \rho_{in}(X)$ on an infinite domain as follows:
\begin{align}\label{eq:step_initial}
    \rho_{in}(X)=
    \begin{cases}
        \rho_0+\rho_1 & \text{for } x < 0\\
        \rho_0 & \text{for } x \ge 0,
    \end{cases}
\end{align}
where $\rho_0=1$ and $\rho_1=3$. We can then directly solve Eq. \eqref{eq:mass_upd} and get the solution of excess density profile $\rho_X(t)-\rho_0= {\cal R}(X/\sqrt{t})$ in terms of a scaling function, ${\cal R}(z)$, where the position is scaled by $z=X/\sqrt{t}$. Moreover, by solving the resultant equation
\begin{align}
    -z\frac{d\mathcal{R}}{dz} = D\frac{d^2{\cal R}}{dz^2}
\end{align}
for ${\cal R}(z)$ with boundary condition $\mathcal{R}(z \to \infty)=0$ and $\mathcal{R}(z \to -\infty)=\rho_1$, we immediately obtain the exact scaling solution,
\begin{align}
    \mathcal{R}(z) = \frac{\rho_1}{2}\left[1-\text{erf}\left(\frac{z}{\sqrt{4D}}\right)\right]\equiv \frac{\rho_1}{2}\text{erfc}\left(\frac{z}{\sqrt{4D}}\right),
    \label{eq:scaled_prof}
\end{align}
 where the scaling variable is defined as $z=x/\sqrt{t}$,  and the complementary error function is given by $\text{erfc}(z)=({2}/{\sqrt{\pi}}) \int_{z}^\infty e^{-t^2}dt$.

 \begin{figure*}
    \centering
     \includegraphics[width=0.5\linewidth]{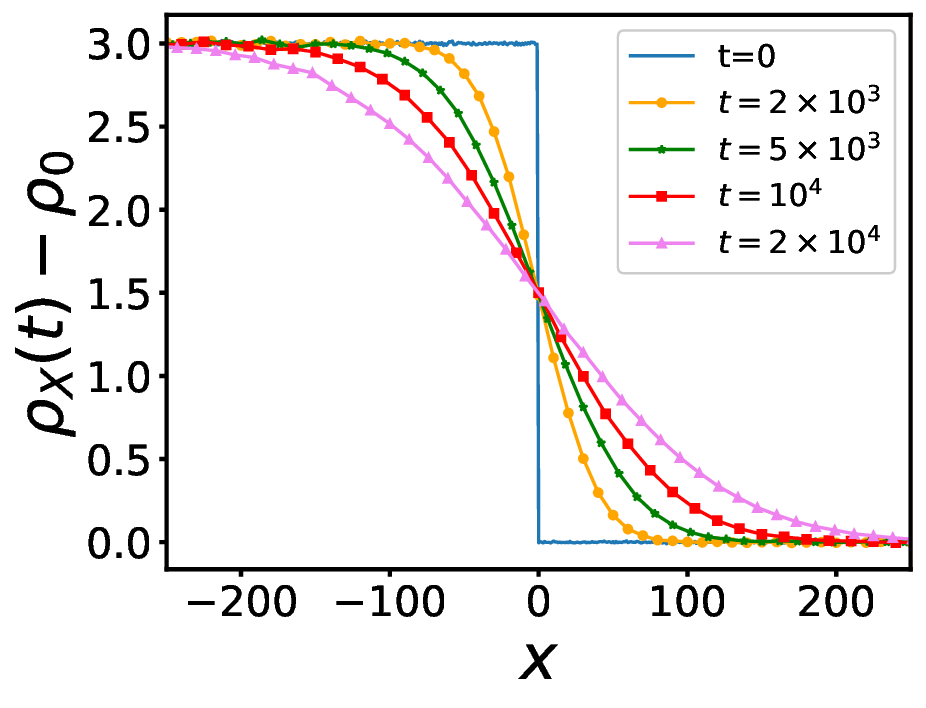}
    \put(-205,160){\textbf{(a)}}
    \includegraphics[width=0.5\linewidth]{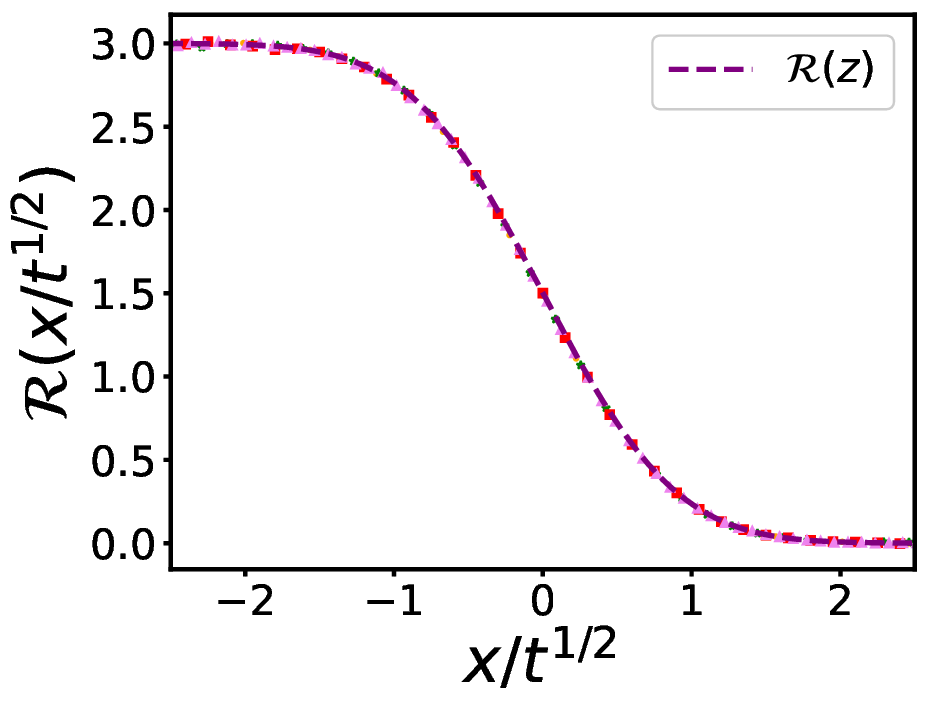}
    \put(-205,160){\textbf{(b)}}
    \caption{  {\it MCM-CoMC I in one dimension.} Panel (a) shows the excess density $(\rho_X(t) - \rho_0)$ of a step initial profile density, plotted against position X at four times: $ t = 2000$ (orange square), $5000$ (green asterisks), $10000 $(red square), and $20000$ (violet triangle). The background density is $\rho_0 = 1$, and the initial density profile height is $\rho_1 = 3$. Panel (b): The scaled shifted density profile $R(z)$ is plotted as a function of the scaling variable $ z = x/\sqrt{t}$. Points represent simulation data, while the purple dashed line is the analytic solution [see Eq. \eqref{eq:scaled_prof}].  }
    \label{fig:step_rlx}
\end{figure*}
In Fig. \ref{fig:step_rlx}, the excess density $(\rho_X(t) - \rho_0)$ of a step initial profile is plotted against position $x$ in Panel (a) at various times: $ t = 2000$ (orange square), $5000$ (green asterisks), $10000 $(red square), and $20000$ (violet triangle). The background density $\rho_0=1$, and the initial density profile height $\rho_1 = 3$. In Panel (b), the scaled shifted density profile $\mathcal{R}(z)$ is plotted as a function of the scaling variable $z = x/\sqrt{t}$, with simulation data points matching closely with the analytic solution (purple dashed line), mentioned in Eq. \eqref{eq:scaled_prof}. One can see that, despite the constrained dynamics due to the additional (CoM) conservation law, the density relaxation in the MCM remains indeed diffusive.

\subsection{Dynamic correlations: One dimension}\label{Sec:1D_MCM_COM_I}

In this section, we calculate various time-dependent quantities in the steady state, including the variance of time-integrated bond current in a finite time interval and dynamic correlations for instantaneous current in one-dimensional MCM-CoMC I. Notably, the global density $\rho$ and the center of mass are conserved in this model, making it particularly interesting to compute these time-dependent quantities exactly. We denote the dynamical correlation function for two observables $A$ and $B$ as $C^{AB}_{r}(t,t^\prime)= \langle A^{i}(t)B^{i+r}(t^\prime) \rangle_c$. The Fourier transform of the dynamical correlation function is written as
\begin{align}
    \tilde{C}_q^{AB}=\sum_rC_r^{AB}e^{iqr},
\end{align}
and then the inverse Fourier transform is given by
\begin{align}
    C_r^{AB} =\frac{1}{L}\sum_q \tilde{C}_q^{AB} e^{-iqr}.
\end{align}

More specifically, we are interested in characterizing time-integrated bond current, $\mathcal{Q}_i(T)$, which is defined as the net amount of mass that flows through a bond $(i,i+1)$ in a time interval $T$ in the steady state.
That is, if mass $\delta m$ is transferred from site $i$ (or, $i+1$) to site $i+1$ ($i$) during an infinitesimal time interval $(t,t+\delta t)$, we measure a positive (negative) instantaneous current $+\delta m/\delta t$ ($-\delta m/ \delta t$) flown through the bond; in other words, the cumulative current increases (decreases) by $+\delta m$ ($-\delta m$) in that time interval $\delta t$. 
The time-evolution equation for the first moment of the cumulative bond current from the following infinitesimal-time update rules,
\begin{align}\label{eq:current_update}
\cq_i(t+dt)=
    \begin{cases}
        \textbf{event}         & \textbf{prob.} \\
        \cq_i(t)+\frac{\tilde{\xi}_im_i(t)}{2}      & dt\\
        \cq_i(t)-\frac{\tilde{\xi}_{i+1}m_{i+1}(t)}{2}    & dt\\
        \cq_i(t)  & (1-2dt),
        \end{cases}
\end{align}
from which we obtain the following equation,
\begin{align}\label{eq:dqdt_1d_com1}
    \frac{d \langle\cq_i(t)\rangle}{dt}\equiv \langle \cj_i(t) \rangle = \frac{\mu_1}{2} \left\langle m_i(t) -m_{i+1}(t)\right\rangle,
\end{align}
where we define instantaneous bond current as $\cj_i(t) \equiv d \cq_i / dt$. that is, the cumulative (time-integrated) bond current in an infinitesimal time interval $\delta t$ is given by $\cq_i(\delta t) = \lim_{\delta t \to 0} \int_t^{t + \delta t} \cj_i(t) dt$. The instantaneous bond current $\cj_i(t) = J^{D}(t) + {\cal J}^{fl}(t)$ can be decomposed into two parts: The diffusive current 
\begin{equation}
\label{JD}
    J^{D}(t) \equiv \frac{\mu_1}{2} [m_i(t) - m_{i+1}(t)]
\end{equation} and fluctuating (or, ``noise'') current ${\cal J}^{fl}(t)$; it is evident that $\langle {\cal J}^{fl} \rangle = 0$. Note that Eq. (\ref{eq:dqdt_1d_com1}) is linear in local masses; this particular {\it linearity property} \cite{Hazra2024Aug} is actually required for obtaining an exact solution to the current statistic in the class of mass transport processes considered in this paper.
We can now write the time-evolution equation for the unequal-time $(t > t')$ current correlation function as
\begin{align}
\label{eq:dqqrttp_1dCOMI}
    \frac{d}{dt}C_r^{\cq\cq}(t,t^\prime) = \frac{\mu_1}{2}\left[C_r^{m\cq}(t,t^\prime)-C_{r-1}^{m\cq}(t,t^\prime)\right].
\end{align}
The above equation has a quite similar structure as derived in Ref. \cite{Hazra2024Aug} for MCMs {\it without} CoM conservation. However, as discussed in the rest of the paper, the explicit solutions to the above equation for MCMs with CoM conservation is qualitatively different from those without the CoM conservation. 
Now proceeding along the lines of Ref. \cite{Hazra2024Aug}, we take Fourier transform of both sides of the above equation and obtain 
\begin{align}\label{eq:qqttpq}
    \frac{d}{dt} \tilde{C}_q^{\cq\cq}(t,t^\prime) = \mu_1(1-e^{iq}) \frac{\tilde{C}_q^{m\cq}(t,t^\prime)}{2}.
\end{align}
To solve Eq. \eqref{eq:dqqrttp_1dCOMI}, we first need to calculate the unequal-time mass-current correlation function $C_r^{m\cq}(t,t^\prime)$, which, in Fourier space, satisfies the following equation,
\begin{align}
\label{eq:cmqttp}
    \frac{d}{dt}\tilde{C}_q^{m\cq}(t,t^\prime) = -\frac{\mu_1}{2}\lambda_q\tilde{C}_q^{m\cq}(t,t^\prime),
\end{align}
where 
\begin{align}
    \lambda_q = 2(1-\cos q),
\end{align}
with the wave number $q=2\pi n/L$ and $n=1, 2, \cdots, L-1$. 
We also require to calculate the equal-time current-current spatial correlation function, which in Fourier space can be written as
\begin{align}
\label{eq:cqqtt}
    \frac{d}{dt}\tilde{C}_q^{\cq \cq}(t,t)&=\mu_1(1-e^{iq})\tilde{C}_q^{m\cq}(t,t)+\tilde{\Gamma}_q,
\end{align}
where $\tilde{\Gamma}_q = \mu_2\langle m^2 \rangle \lambda_q /4$ is related to the strength of fluctuating current as given below:
\begin{align}\label{eq:jqjq}
    \langle \tilde{\cj}_q^{(fl)}(t)\tilde{\cj}_{q^\prime}^{(fl)}(t^\prime)\rangle = L \tilde{\Gamma}_q\delta_{q,-q^\prime}\delta(t-t^\prime).
\end{align}
The right-hand side of the above equation indicates that the strength of the fluctuating current is short-ranged (delta correlated) in both spatial and temporal scales \cite{Hazra2024Aug}.
Note that solving Eq. (\ref{eq:cmqttp}) requires calculation of equal-time mass-current spatial correlation, which, in the Fourier space, can be written as
\begin{align}
\label{eq:cmqtt}
    \frac{d}{dt}\tilde{C}_q^{m\cq}(t,t)=-\frac{\mu_1}{2}\lambda_q\tilde{C}_q^{m\cq}(t,t)+\tilde{f}_q.
\end{align}
Here the source term $\tilde{f}_q$ can be written in terms of the steady-state density correlation $\tilde{C}_q^{mm}$ as
\begin{align}
\label{eq:f_q}
   \tilde{f}_q=(1-e^{-iq})\left[D\tilde{C}^{mm}_q-\frac{\mu_2\langle m^2\rangle}{4}\lambda_q\right].
\end{align}

\subsubsection{Density correlation and structure factor.} 

It is quite convenient, and instructive, to express the steady-state density correlation function in Fourier mode $\tilde{C}^{mm}_q$. To this end, we write the time evolution of the Fourier mode for equal-time mass-mass correlation function as 
\begin{equation}
    \frac{d}{dt}\tilde{C}_q^{mm}(t,t) = -2D\lambda_q\tilde{C}_q^{mm}(t,t)+B(q),
\end{equation}
where the quantity $B(q)=\mu_2 \langle m^2 \rangle\lambda_q^2/4$ depends on the second moment of onsite mass. We readily obtain the solution to the above equation, which is given by 
\begin{align}
    \tilde{C}^{mm}_q(t,t) = \left(1-e^{-2D\lambda_qt}\right)\frac{\langle m^2\rangle\mu_2}{8D}\lambda_q,
\end{align}
which, in the steady state (in the limit $t\to\infty$), leads to the static density correlation function in the Fourier space,
\begin{equation}
\label{eq:c_qmm}
    \tilde{C}_q^{mm} = \frac{\mu_2\rho^2}{2(2\mu_1-\mu_2)} \lambda_q.
\end{equation}
Now onward, we follow the convention that, if the time argument in the correlation function is not explicitly mentioned, e.g., $\tilde{C}_q^{mm}$ - a function of density and other parameters, it would simply imply that the correlation function is calculated in the steady state and thus does not depend on time. Note that, in this case, the above-mentioned steady-state density-correlation function $\tilde{C}^{mm}_q$ is related to the strength of fluctuating current as 
\begin{align}\label{eq:lambda_q_cq}
    \tilde{\Gamma_q}=2D\tilde{C}^{mm}_q.
\end{align}
Now by taking inverse Fourier transform, Eq. \eqref{eq:c_qmm} leads to the correlation function in real space as given below:
\begin{align}
\label{eq:c_rmm}
    C_r^{mm}(\rho) = \frac{\mu_2\rho^2}{2(2\mu_1-\mu_2)}\left[ 2\delta_{r,0}-(\delta_{r,1}+\delta_{r,-1}) \right],
\end{align}
which is strictly finite-ranged and valid for any finite $L$. 
Moreover, Eq. \eqref{eq:lambda_q_cq} leads to a fluctuation relation between the two-point spatial correlation function for mass and the intensity of fluctuating current, 
\begin{align}\label{eq:gamr_cr}
    C_r^{mm}(\rho) = \frac{\Gamma_r(\rho)}{2D}.
\end{align}
Now, to calculate the dynamic structure factor, we define Fourier transform $\delta \tilde{m}_q(t)$ of excess mass $\delta m_i (t) = (m_i(t) - \rho)$ at site $i$, which satisfies the following time-evolution equation for the Fourier modes $\delta \tilde{m}_q(t)$,
\begin{align}
	\label{eq:delm_q}
	\frac{\partial}{\partial t} {\delta \tilde m}_q(t) = -D\lambda_q \tilde{\delta m}_q(t) + (e^{iq} - 1)\tilde{\mathcal{J}}^{(fl)}_q(t).
\end{align}
In the above equation, the first term in the right hand side arises from diffusion and the second term is due to the fluctuating (or, the ``noise'') current. By using Eqs. (\ref{eq:delm_q}) and (\ref{eq:jqjq}) and performing some algebraic manipulations, we obtain the dynamic structure factor,
\begin{align}\label{eq:struc_dynamical_1D_I}
    S(q,t) = \frac{\langle |\delta {\tilde m}_q(t)|^2 \rangle}{L\rho} &= \frac{1}{\rho}\tilde{C}^{mm}_q(t,t)\\ \nonumber &= \frac{1}{\rho}\left(1-e^{-2D\lambda_q t}\right)\tilde{C}^{mm}_q ,
\end{align}
where $\tilde{C}^{mm}_q(t,t)$ and $\tilde{C}^{mm}_q \equiv \lim_{t \to \infty} \tilde{C}^{mm}_q(t,t)$ are the steady-state dynamic and static density-correlation functions, respectively. Exact calculation of dynamic structure factor in a driven many-body system is quite challenging in general and has been done before only in a few cases in the past, e.g., in Refs. \cite{Sachdeva2014Feb, Karevski2017Jan}. The steady-state static structure factor $S(q) = \lim_{t \to \infty} S(q, t)$ is then given by
\begin{equation}
\label{eq:struc_st_1D_I}
    S(q) = \frac{1}{\rho} \tilde{C}^{mm}_q = \frac{\mu_2\rho}{2(2\mu_1-\mu_2)} \lambda_q \simeq \frac{\mu_2\rho}{2(2\mu_1-\mu_2)} q^2,
\end{equation}
which vanishes as $q \to 0$; here, in the last step of the above expression, we have used  a small-$q$ approximation of $\lambda_q \approx q^2$. Indeed, this particular $q$-dependence of the structure factor, i.e., $S(q) \sim q^2$ in the limit of small $q$, implies a ``Class-I'' hyperuniformity, implying an extreme suppression of density fluctuations in spatial domain, as categorized in Ref. \cite{Torquato2018Jun}.

\begin{figure*}
    \centering
    \includegraphics[width=0.5\linewidth]{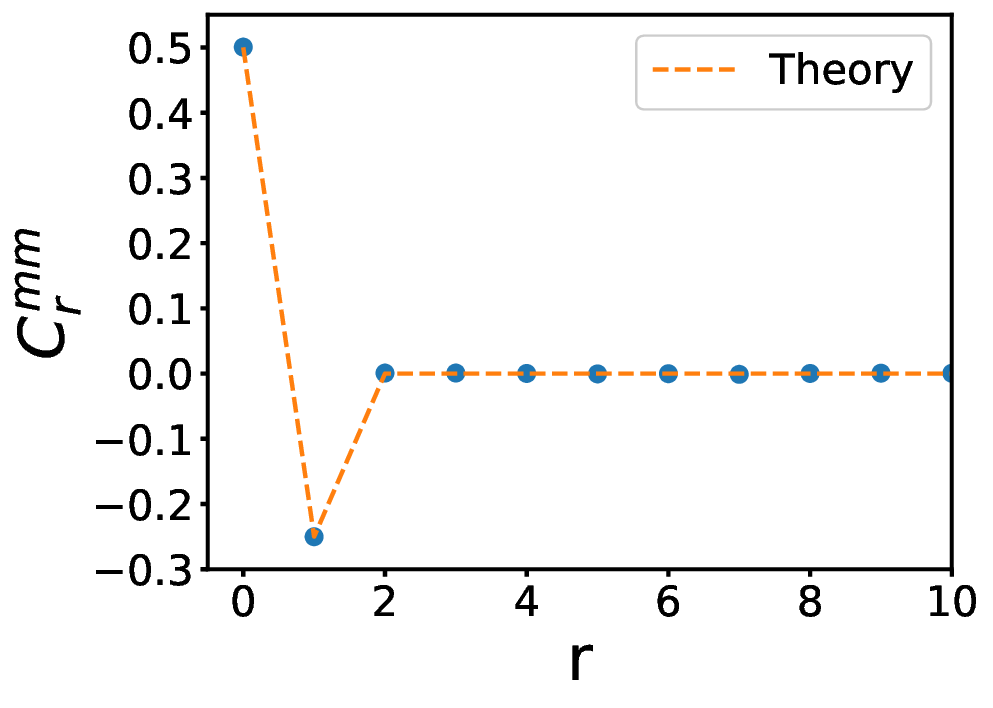}
    \put(-110,170){\textbf{(a)}}
    \includegraphics[width=0.5\linewidth]{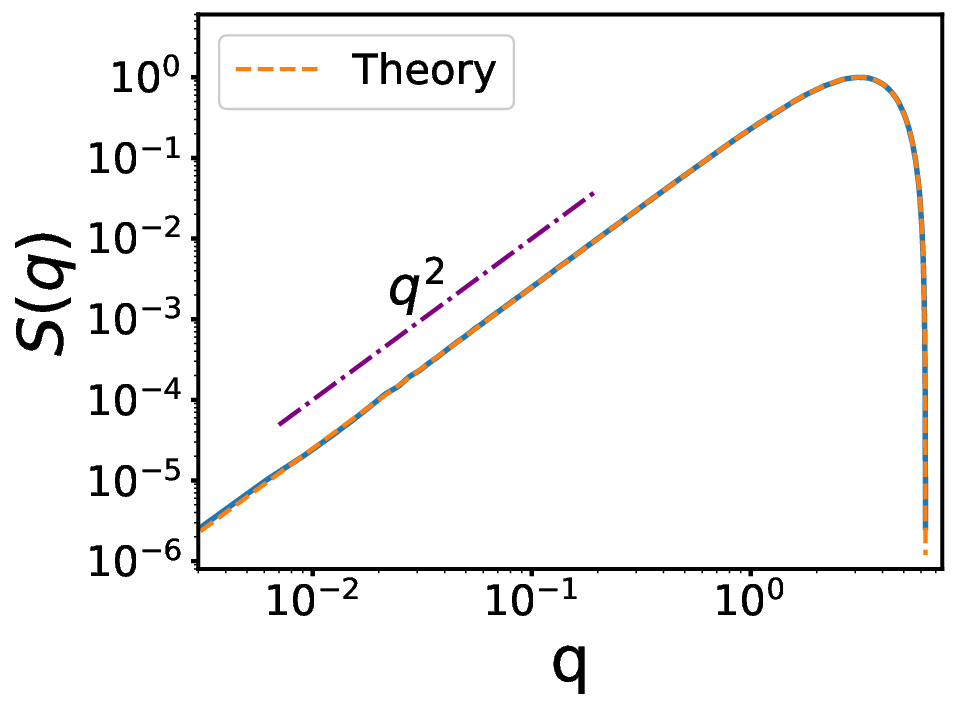}
    \put(-110,170){ \textbf{(b)}}
    \caption{ {\it MCM-CoMC I in one dimension.} Panel (a): We plot steady-state density correlation function, $C_r^{mm}$, as a function of spatial distance $r$. Panel (b): Steady-state structure factor $S(q)$ is plotted as a function of wave number $q$. The purple dashed-dotted line represent the asymptotic behavior, which scales as $q^2$ for small $q$. In both the panels (a) and (b), the orange dashed lines represent theoretical predictions obtained from Eqs. \eqref{eq:c_rmm} and \eqref{eq:struc_st_1D_I}, respectively. Simulations are performed for global density $\rho=1.0$ and system size $L=1000$. }
    \label{fig:cor_stuc_COMI_1D}
\end{figure*}

In panel (a) of Fig. \ref{fig:cor_stuc_COMI_1D}, we have plotted the steady-state density correlation function $C_r^{mm}$ as a function of spatial distance $r$. In panel (b) of the same figure, we have plotted the static structure factor $S(q)$ as a function of wave number $q$. In both panels (a) and (b), the orange dashed lines represent theoretical predictions, which are obtained from Eqs. \eqref{eq:c_rmm} and \eqref{eq:struc_st_1D_I}, respectively. Simulations are performed for global density $\rho=1.0$ and system size $L=1000$, and match quite well with theory.

\subsubsection{Calculation of current correlation functions.}

To obtain unequal-time current-current correlation function, we first evaluate equal-time current-current and mass-current correlation functions from Eqs. \eqref{eq:cqqtt} and \eqref{eq:cmqtt}, respectively.  Next, we calculate the unequal time mass-current correlation using Eq. \eqref{eq:cmqttp}. Finally, we derive the full solution for the unequal time current-current correlation from Eq. \eqref{eq:dqqrttp_1dCOMI} in the Fourier representation,
\begin{align}\label{eq:cqQQttp_MCM_com_1D}
    \tilde{C}_q^{\cq\cq}(t,t^\prime)= \tilde{\Gamma}_q t^\prime &-\frac{\mu_2(D\lambda_q\rho)^2}{(2\mu_1-\mu_2)}\Bigg[\int\limits_0^{t^\prime}dt^{\prime\prime}\int\limits_{0}^{t^{\prime\prime}}dt^{\prime\prime\prime}e^{-D\lambda_q(t^{\prime\prime}-t^{\prime\prime\prime})}\\ \nonumber &+\frac{1}{2}\int\limits^t_{t^\prime}dt^{\prime\prime}\int\limits_{0}^{t^{\prime\prime}}dt^{\prime\prime\prime}e^{-D\lambda_q(t^{\prime\prime}-t^{\prime\prime\prime})} \Bigg].
\end{align}
 The equal-time bond-current correlation $C_0^{\cq\cq}(T,T) \equiv \langle \cq_i^2(T)\rangle_c$, or the variance (the second cumulant), can now be obtained by first taking the inverse Fourier transform in Eq.(\ref{eq:cqQQttp_MCM_com_1D}) and then setting $t=t^\prime=T$ and $r=0$, as follows:
\begin{align}
\label{eq:Qi2t}
    \langle \cq_i^2(T)\rangle_c &= \langle \cq_i^2(T)\rangle = T \left [\Gamma_0-DC_0^{mm}\frac{1}{L}\sum\limits_{q\ne0}\lambda_q \right]\\ \nonumber &+\frac{\mu_2\rho^2}{(2\mu_1-\mu_2)}\frac{1}{L}\sum\limits_{q\ne0}\left( 1-e^{-\frac{\mu_1}{2}\lambda_qT}\right).
\end{align}
In the long-time regime, by collecting leading-order terms, Eq. \eqref{eq:Qi2t} can be written in the following asymptotic form,
\begin{align}
\langle \cq_i^2(T) \rangle \simeq  A_1 T + A_2 + A_3 T^{-1/2},
\end{align}
where $A_1$, $A_2$ and $A_3$ are constant (time-independent) coefficients, which in general depend on density and other model parameters.
Quite remarkably, in this case (presumably, generically in one-dimensional models as discussed in the next section), we find that the coefficient corresponding to the first (linear-time-growth) term in the above equation vanishes, i.e., 
\begin{align}
\label{A1-zero-1D}
    A_1 = \left [\Gamma_0-DC_0^{mm}\frac{1}{L}\sum\limits_{q}\lambda_q \right] = 0,
\end{align}
obtained by using the identity $\sum_{q}\lambda_q=2L$ (note that $\lambda_0=0$) and the fluctuation relation $\Gamma_0=2DC_0^{mm}$ from Eq. \eqref{eq:gamr_cr}.
Now, in the thermodynamic limit, with $L \to \infty$ and density being fixed,  Eq. \eqref{eq:Qi2t} can be written in an integral form as given below:
\begin{align}\label{eq:c_0ttMCM_1D_int}
    C_{0}^{\cq\cq}(T,T) &\simeq \frac{\mu_2\rho^2}{(2\mu_1-\mu_2)}\frac{1}{\pi} \int\limits_0^\pi dq\left( 1-e^{-\frac{\mu_1}{2}\lambda (q)T}\right)\\ \nonumber
    &=\frac{\mu_2\rho^2}{(2\mu_1-\mu_2)}\left[1-e^{-\mu_1T}I_0(\mu_1T)\right],
\end{align}
where $I_0(x)=(1/\pi)\int_0^\pi e^{x\cos q}dq$ is the modified Bessel function of the first kind and we denote $\lambda(q)=2(1-\cos q)$.
By performing an asymptotic analysis, Eq. \eqref{eq:c_0ttMCM_1D_int} can be written for large time $T$ as follows:
\begin{align}
\label{eq:assym_qi2t}
    \langle \cq_i^2(T)\rangle \simeq \frac{\mu_2\rho^2}{(2\mu_1-\mu_2)} - \frac{\mu_2\rho^2}{(2\mu_1 -\mu_2)\sqrt{2\pi\mu_1}} T^{-{1}/{2}},
\end{align}
where we have used the approximation $\lambda(q)\approx q^2$.
That is, in the thermodynamic limit, the variance of time-integrated bond current in a long time interval $T \to \infty$ saturates to a finite value,
\begin{align}
\label{eq:saturation}
     A_2(\rho) = \frac{\mu_2\rho^2}{(2\mu_1-\mu_2)},
\end{align}
through a power-law decay $T^{-1/2}$.
The third coefficient $A_3$ can also be identified from Eq. \eqref{eq:assym_qi2t} as
\begin{align}
    A_3(\rho) = - \frac{\mu_2\rho^2}{(2\mu_1 -\mu_2)\sqrt{2\pi\mu_1}}.
\end{align}

The fact that the coefficient $A_1$ vanishes here implies that, in the long-time limit, the variance ${\cal Q}_i^2(T)$ of time-integrated bond current in one-dimensional MCM {\it with} CoM conservation saturates as a function of time $T$ as opposed to growing as $T^{1/2}$ in one-dimensional MCMs {\it without} CoM conservation \cite{Hazra2024Aug}. 
 Physically, in the former case, it implies a stronger anti-correlations between currents developing at two different times; it could be understood on the physical ground that, whenever there is a current generated at some point, there is also a reverse current of the same magnitude generated in the opposite direction.
This extreme suppression of dynamic current fluctuation is somewhat analogous to extreme suppression of dynamic activity fluctuations previously observed in a model of sandpiles, called the Oslo model, in Ref. \cite{Garcia-Millan2018Jul} (the hyperuniformity exponent $b=0$ in both cases). 
Furthermore, the expression in Eq. \eqref{eq:assym_qi2t} has a quite similar structure to that obtained previously, albeit through an approximate closure scheme, for the one dimensional  Oslo model in the far-from-critical regime \cite{Mukherjee2024Aug}. 
However, unlike in the Oslo model, the calculation scheme for MCMs in this paper is exact and the corresponding coefficients $A_1$, $A_2$ and $A_3$ are obtained explicitly as a function of density and the other parameters.

\subsubsection{Instantaneous bond-current correlation functions.}  

\begin{figure*}
    \centering
    \includegraphics[width=0.33\linewidth]{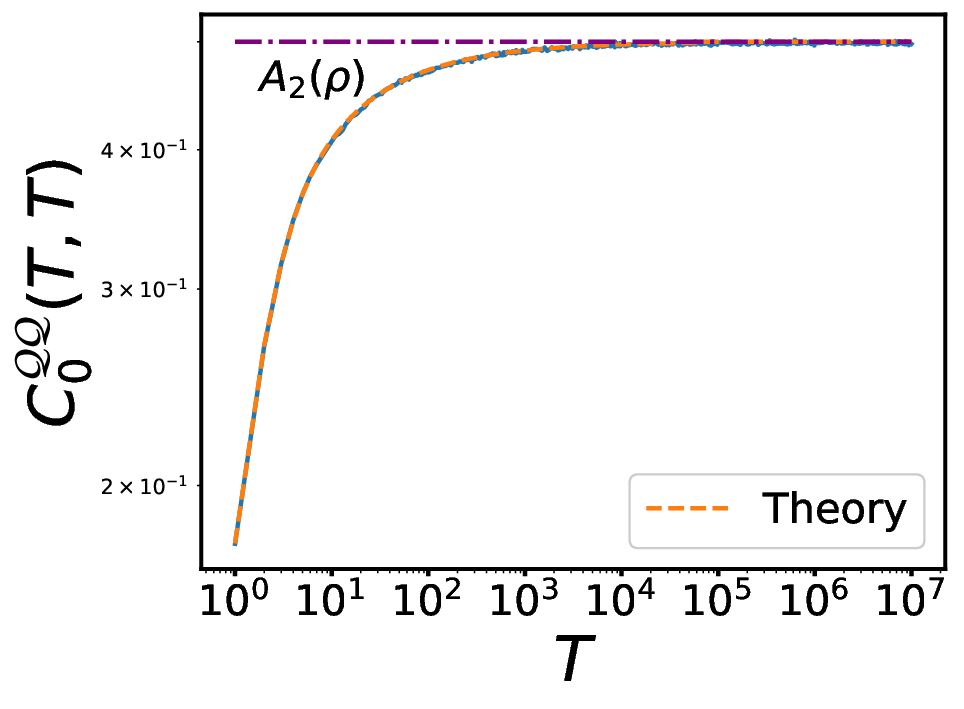}
    \put(-70,110){\textbf{(a)}}
    \includegraphics[width=0.33\linewidth]{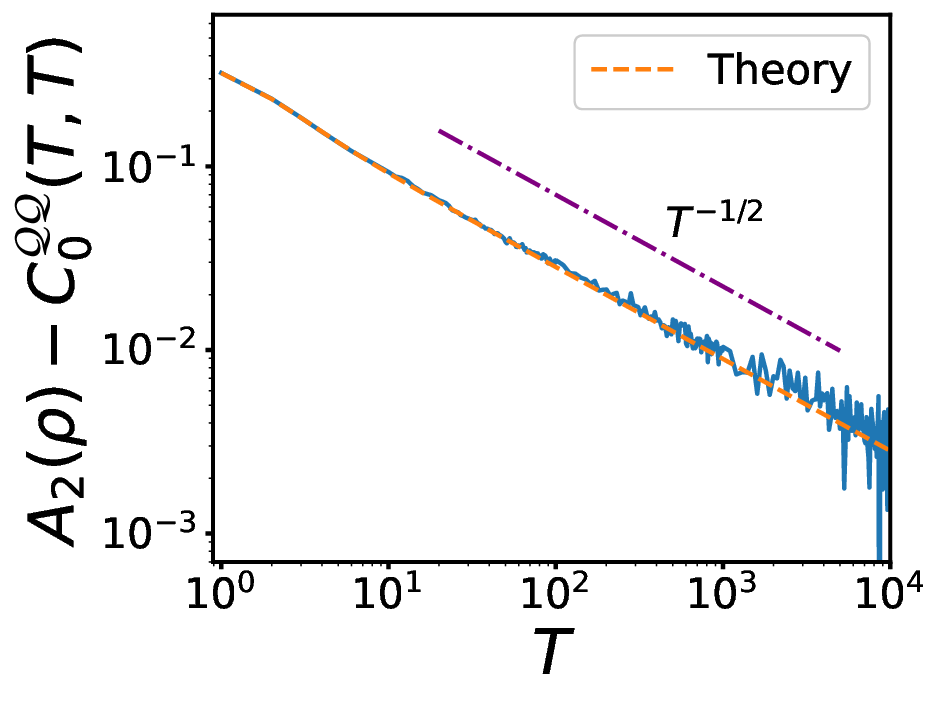}
    \put(-85,110){ \textbf{(b)}}
    \includegraphics[width=0.33\linewidth]{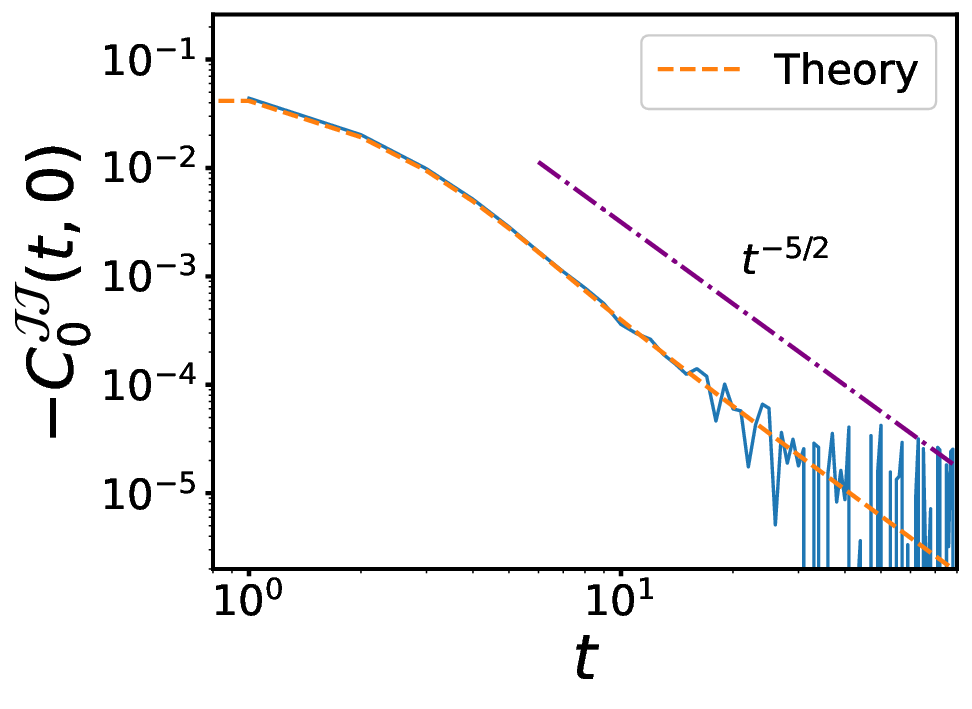}
    \put(-70,110){\textbf{(c)}}
    \caption{ {\it MCM-COM-I in one dimension.} Panel (a): Time-integrated bond-current fluctuation or variance $\langle \cq_i^2(T)\rangle=C_0^{\cq\cq}(T,T)$ is plotted as a function of time $T$. Theory line (orange dashed), obtained from Eq. \eqref{eq:c_0ttMCM_1D_int}, agrees quite well with simulation perfectly. The guiding line (purple dashed) indicates the saturation value obtained from Eq.\eqref{eq:saturation}. Panel (b): The relative time-integrated bond-current fluctuation, $A_2(\rho) - C_0^{\cq\cq}(T,T)$, is plotted as a function of time $T$ for both simulation (blue) and theory (orange). The asymptotic $T^{-1/2}$ (purple) is obtained from Eq.(\ref{eq:assym_qi2t}). Panel (c), negative of instantaneous bond-current correlation $-C_0^{\cj\cj}(t,0)$ is plotted as a function of $t$. The orange dashed line represents theory as in Eq. \eqref{eq:cjjt_COM1D}. The purple dashed-dotted line represents the asymptotic power-law decay $t^{-5/2}$, obtained from theory Eq. \eqref{eq:cjjtasym_com1d}. In all panels, simulation data (blue line) is for system size $L = 1000$ and global density $\rho = 1$. }
    \label{fig:MCM_COM_I_1D_dync}
\end{figure*}

We now calculate the temporal correlation of instantaneous current correlations, $C_r^{\cj\cj}(t,0)$ for $t \ge 0$, which can be expressed as:
\begin{align}\label{eq:crjjt_1D_com}
    C_r^{\cj\cj}(t,0)&=\left[\frac{d}{dt}\frac{d}{dt^\prime}C_{r}^{\cq\cq}(t,t^\prime)\right]_{t^\prime=0, t\ge0}\\ \nonumber &= \Gamma_r\delta(t)-\frac{\mu_1^2\mu_2\rho^2}{8(2\mu_1-\mu_2)}\frac{1}{L}\sum_q e^{-\frac{\mu_1}{2}\lambda_q t}\lambda_q^2e^{-iqr}.
\end{align}
In the large system size limit, the above equation with $r=0$ can be written in the following form:
\begin{align}\label{eq:cjjt_COM1D}
    C_0^{\cj\cj}(t,0) \simeq -\frac{\mu_1^2\mu_2\rho^2}{8(2\mu_1-\mu_2)}e^{-\mu_1 t}\left[2I_0(\mu_1t)-\frac{I_1(\mu_1 t)}{\mu t}\right],
\end{align}
where we have taken $t>0$ and $I_\nu(z)$ denotes the modified Bessel function of the first kind of order $\nu$. Also, in the long-time limit and for large-system-size limit, the Eq. \eqref{eq:crjjt_1D_com} with $r=0$ has the following asymptotic form,
\begin{align}\label{eq:cjjtasym_com1d}
    C_0^{\cj\cj}(t,0) \simeq -\frac{3\mu_2\rho^2}{16\sqrt{2\pi\mu_1}(2\mu_1-\mu_2)}t^{-5/2}.
\end{align}
In Fig. \ref{fig:MCM_COM_I_1D_dync}, panel (a), the variance $C_{0}^{\cq\cq}(T,T)=\langle \cq_i^2(T)\rangle$ of time-integrated current in a time interval $T$ is plotted as a function of $T$. The theory line (orange dashed), obtained from Eq. \eqref{eq:c_0ttMCM_1D_int}, matches with the simulation perfectly. The guiding line (purple dash-dot) indicates the saturation value obtained from Eq.\eqref{eq:saturation}. Panel (b): The relative time-integrated bond-current fluctuation, $A_2(\rho) - C_{0}^{\cq\cq}(T,T)$, is plotted as a function of time $T$ for both simulation (solid blue line) and theory (orange dashed). The purple dashed-dotted line represents $T^{-1/2}$, as described in Eq.\eqref{eq:assym_qi2t}. Panel (c): Negative of instantaneous bond-current correlation $-C_0^{\cj\cj}(t,0)$ is plotted as a function of $t$. The theory line (orange dashed), mentioned in Eq. \eqref{eq:cjjt_COM1D}, matches the simulation quite well. The asymptotic decay as $t^{-5/2}$ (purple dashed-dotted) is consistent with Eq. \eqref{eq:cjjtasym_com1d}. In all the panels simulation data (blue line) is for system size of $L = 1000$ and a global density $\rho = 1$.

\section{One-dimensional MCM-CoMC II: Finite-range mass transfer}\label{Sec:1D_MCM_COM_II}

In this section, we consider a variant of the CoM-conserving mass chipping model on a one-dimensional periodic ring, now with {\it finite-range} mass-transfer rule, where the chipped-off mass can coalesce with not only the nearest-neighbor masses, but also the next-nearest-neighbor masses. More specifically, mass $m_i$ at site $i$ is chipped off with unit rate and a random fraction $\xi_i$ of mass $m_i$, i.e., $\xi_i m_i$ is retained, while the remaining fraction of mass $(1-\xi_i) m_i$ is fragmented into two equal halves, each having mass $(1-\xi_i) m_i/2$. These two chunks of fragmented masses then coalesce, with equal probability, either with the two nearest-neighbor masses at $i+1$ and $i-1$ or with the next-nearest-neighbor masses at $i+2$ and $i-2$.
It is worth noting that, in a finite-range (say, of range $k$) mass transfer process, when a random fraction of the chipped-off mass, $\delta m$, is transferred from the $i$th site to $(i+k)$th site (where $k=2$ in the model considered in this section) during the time interval $(t, t+\delta t)$, the cumulative currents increase by $\delta m$ across all bonds from $(i, i+1)$ to $(i+k-1, i+k)$. Similarly, if the chipped-off mass, $\delta m$, is transferred from the $i$th site to  $(i-k)$th site, the cumulative currents decrease by $\delta m$ across all bonds from $(i-1, i)$ to $(i-k, i-k+1)$. 

\begin{figure*}
    \centering
    \includegraphics[width=0.5\linewidth]{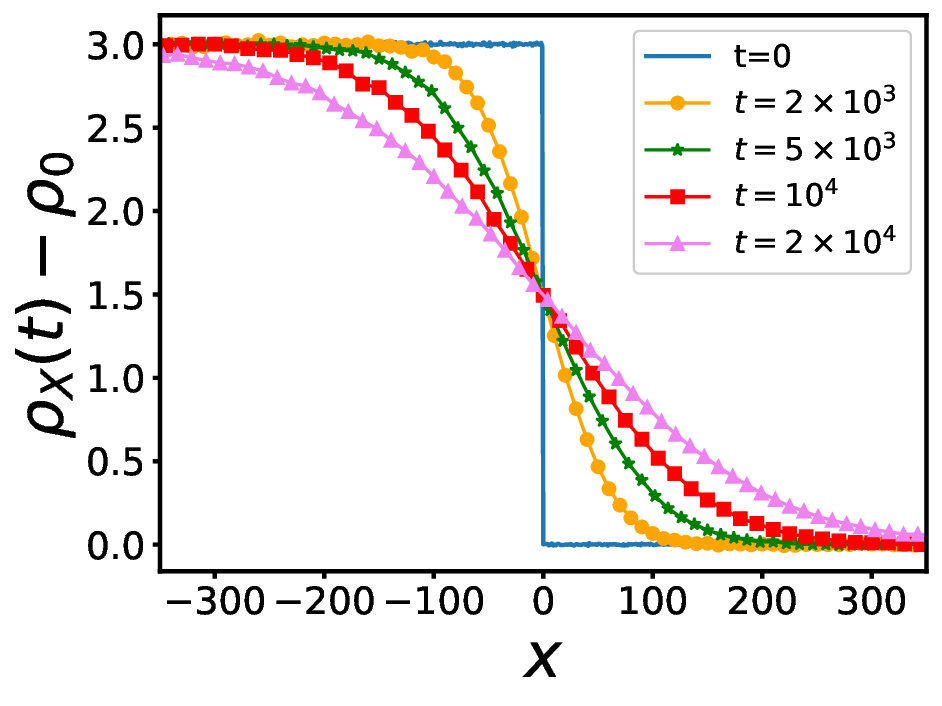}
    \put(-210,160){\textbf{(a)}}
    \includegraphics[width=0.5\linewidth]{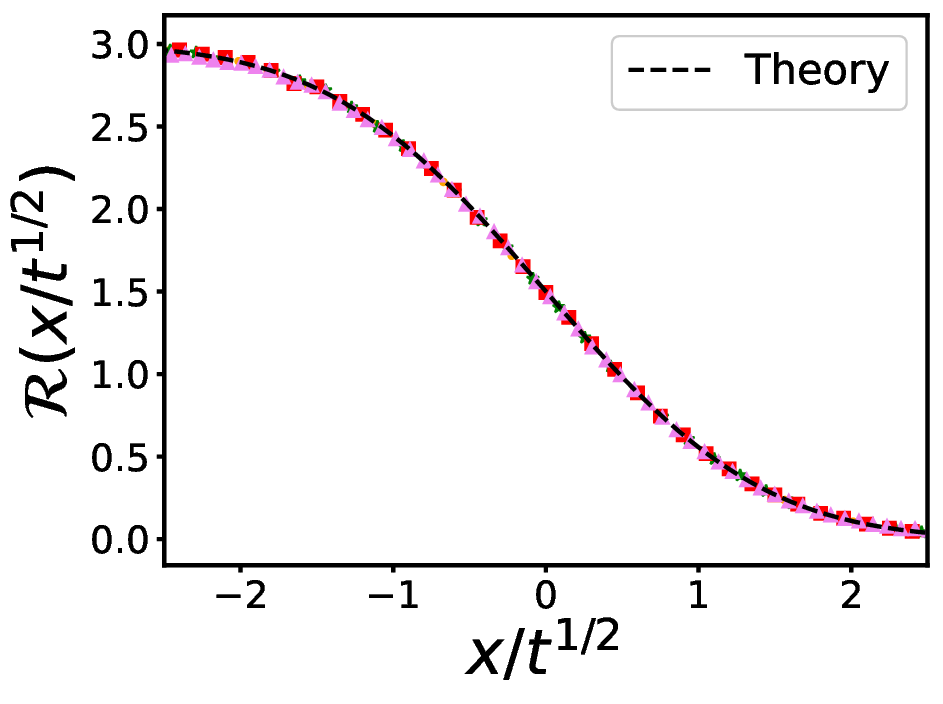}
    \put(-210,160){\textbf{(b)}}
    \caption{ {\it MCM-CoMC II in one dimension.} Panel (a): The excess density $(\rho_X(t) - \rho_0)$ of an initially stepped density profile is plotted against position $X$ at four different times: $t = 2000$ (orange squares), $5000$ (green asterisks), $10000$ (red squares), and $20000$ (violet triangles). The background density is $\rho_0 = 1$, with an initial profile height of $\rho_1 = 3$. Panel (b): The scaled shifted density profile $R(z)$ is plotted against the scaling variable $z = X/t^{1/2}$. Points represent simulation data, while the black dashed line [Eq. \eqref{eq:scaled_prof}] indicates the analytic solution with $D=5\mu_1/4$. }
    \label{fig:Hydro_MCM_COM_1D_I}
\end{figure*}

\begin{figure*}
    \centering
    \includegraphics[width=0.33\linewidth]{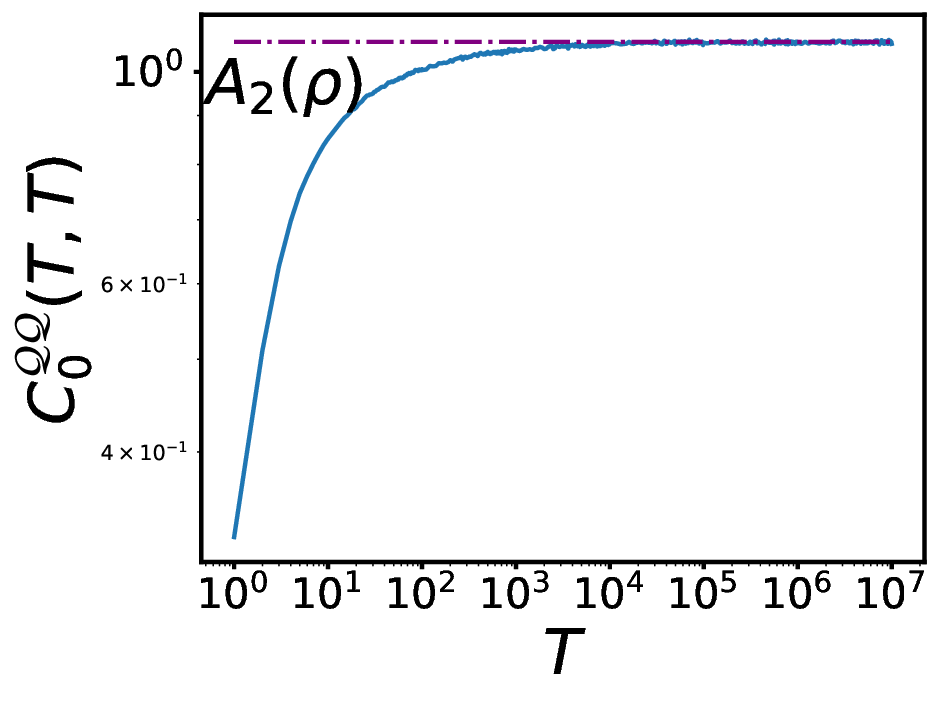}
    \put(-70,110){\textbf{ (a)}}
    \includegraphics[width=0.33\linewidth]{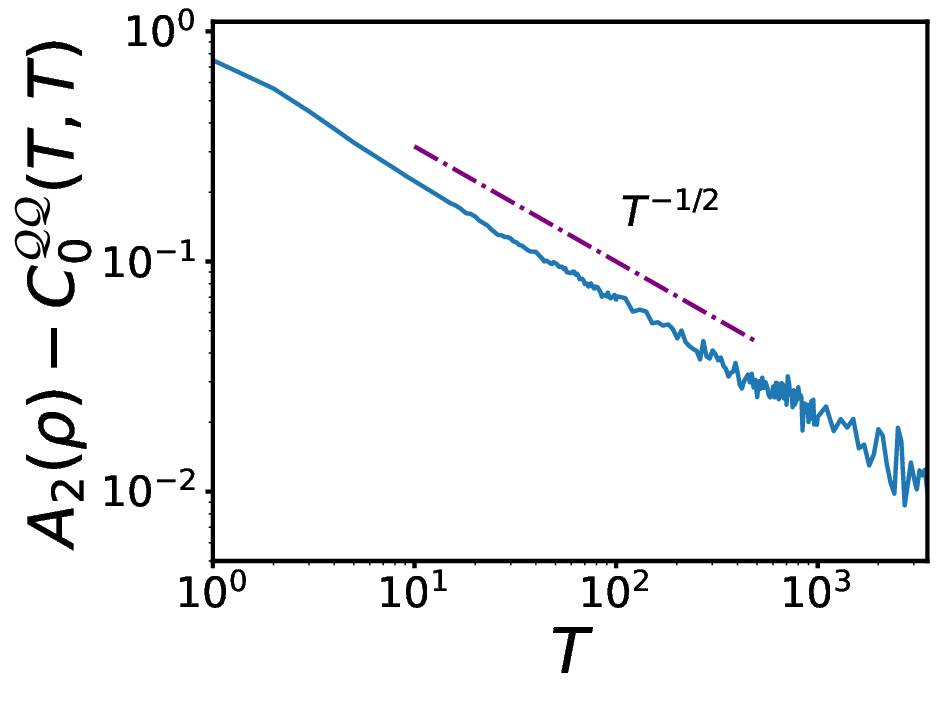}
    \put(-85,110){\textbf{(b)}}
    \includegraphics[width=0.33\linewidth]{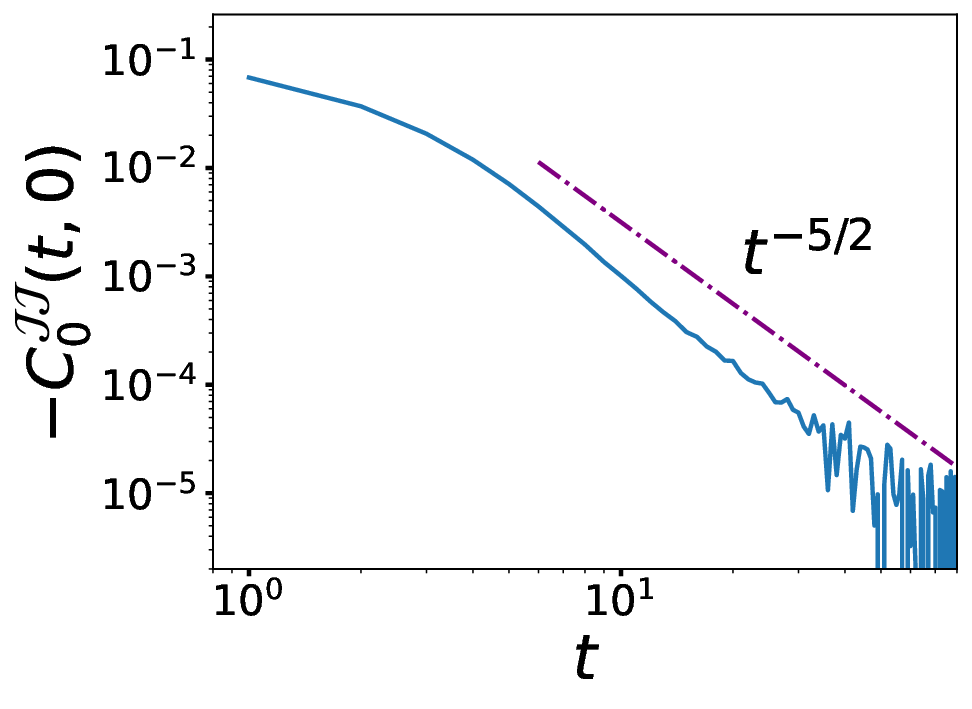}
    \put(-80,110){\textbf{ (c)}}
    \caption{{\it MCM-CoMC II in one dimension.} Panel (a): the time integrated bond-current fluctuation is plotted as a function of time $T$. The purple dashed-dotted line represents the saturation value $A_2(\rho)=1.075$(fitted). Panel (b):  The relative bond-current fluctuation, $A_2(\rho)-C_{0}^{\cq\cq}(T,T)$, is plotted as a function of time $T$. Purple dashed-dotted line is the asymptotic power law decay $T^{-1/2}$, obtained from Eq. \eqref{eq:assym_qi2t}. Panel (c): Negative of instantaneous bond-current correlation $-C^{\cj\cj}_0(t,0)$ is plotted as a function of time $t$. The asymptotic line (purple dashed-dotted) is the theoretical estimation obtained from Eq. \eqref{eq:cjjtasym_com1d}.  In the all the panels, we have taken system size $L=1000$ and global density $\rho=1.0$.}
    \label{fig:cjjt_com_II_1D}
\end{figure*}

We now present simulation results to demonstrate that the qualitative results derived in the previous section are indeed quite robust for one dimensional systems.  In Fig. \ref{fig:Hydro_MCM_COM_1D_I}, panel (a), the excess density profile $(\rho_X(t)-\rho)$ of an initially stepped density profile [see Eq. \eqref{eq:step_initial}] is plotted against position $X$ at four different times: $t = 2000$ (orange squares), $5000$ (green asterisks), $10000$ (red squares), and $20000$ (violet triangles). Panel (b) depicts the scaled density profile ${\cal R}(X/t^{1/2})$ as a function of the scaled position $X/t^{1/2}$. The solution (black dashed line) obtained from Eq. \eqref{eq:scaled_prof} with the theoretically obtained bulk-diffusion coefficient $D=5\mu_1/4$ (different from the model described in the previous section)  agrees quite well with the simulation data.
In panel (a) of Fig. \ref{fig:cjjt_com_II_1D}, the variance $C_0^{\cq\cq}(T,T)$ of time-integrated bond current is plotted as a function of time $T$. In panel (b) of the same figure,  the relative bond-current fluctuation, $A_2(\rho)-C_{0}^{\cq\cq}(T,T)$, is plotted as a function of time $T$. Purple dashed-dotted line is the asymptotic power-law $T^{-1/2}$ as derived in Eq. \eqref{eq:assym_qi2t}. In panel (c), negative of instantaneous bond current correlation, $-C^{\cj\cj}_0(t,0)$, is plotted as a function of time $t$. In the all panels, we have taken system size $L=1000$ and global density $\rho=1.0$.

\section{Two-dimensional MCM-CoMC I}\label{Sec:2D_MCM_COM_I}

\subsection{Model and definition of current}

In this section, we introduce a model defined on a two-dimensional periodic square lattice with the following dynamical rules; for simplicity, we consider only nearest-neighbor mass transfer. Here, a site $(i,j)$ retains a random fraction of its mass $\xi m_{i,j}$, and the remaining fraction of mass, $(1-\xi) m_{i,j}$, is divided equally into four parts, each having $(1-\xi) m_{i,j}/4$, which is then transferred to each of the four nearest neighbors $(i+1,j)$, $(i-1,j)$, $(i,j+1)$, and $(i,j-1)$.

In a two-dimensional square lattice, the bond current is defined similar to that in the one-dimensional case. The time-integrated bond current along the $x$ direction, denoted as $\cq_x^{i,j}(T)$, represents the net flux of mass between site $(i, j)$ and site $(i+1, j)$ in the time interval $T$: If an amount of mass $\delta m$ is transferred from site $i$ to $i+1$ (or, $i+1$ to $i$) during an time interval, say $\delta t$, $\cq_x^{i,j}$ increases (or, decreases) by $\delta m$, i.e., $\cq_x^{i,j} \rightarrow \cq_x^{i,j} + \delta m$ (or, $\cq_x^{i,j} \rightarrow \cq_x^{i,j} - \delta m$). Similarly, the time-integrated bond current along the $y$ direction, denoted as $\cq_y^{i,j}(T)$, represents the net flux of mass through the bond between sites $(i, j)$ and $(i, j+1)$ during the same interval $T$. The time-integrated bond current $\cq_\alpha^{i,j}(T)$, with $\alpha \in {x, y}$, can be expressed in terms of the instantaneous bond current $\cj_\alpha^{i,j}(t)$ as $\cq_\alpha^{i,j}(T)=\int_t^{t+T} \cj_\alpha^{i,j}(t') dt'$, with $\alpha\in x,y$.
In the schematic diagram Fig. \ref{fig:curr_def}, a mass of amount $4 \delta m$ is chipped off from site $(i, j)$ and fragmented into $4$ equal parts with each of the part $\delta m$ being transferred to each of its four neighboring sites during the time interval $(t, t+\delta t)$. As a result, along the $x$ axis, a mass $\delta m$ is transferred from site $(i, j)$ to site $(i+1, j)$, and from site $(i, j)$ to site $(i-1, j)$, thus generating a positive instantaneous current $\cj_x^{i,j} = +\delta m/\delta t$  through the bond between sites $(i, j)$ and $(i+1, j)$ and a negative instantaneous current $\cj_x^{i-1,j} = -\delta m/\delta t$ through the bond between sites $(i-1, j)$ and $(i, j)$. Similarly, along the $y$ axis, a mass $\delta m$ is transferred from site $(i, j)$ to site $(i, j+1)$ and from site $(i, j)$ to site $(i, j-1)$), generating a positive instantaneous current $\cj_y^{i,j} = +\delta m/\delta t$ through the bond $(i, j)$ and $(i, j+1)$ and a negative instantaneous current $\cj_y^{i,j-1} = -\delta m/\delta t$  through the bond $(i, j-1)$ and $(i, j)$.

\begin{figure}
    \centering
    \begin{tikzpicture}
    \def\rows{2}
    \def\cols{2}
    \def\spacing{3} 
    \def\del{0.5}
    \foreach \x in {0,...,\cols} {
        \draw[dotted, thick] (\spacing*\x,-\del)--(\spacing*\x,\spacing*\rows+\del);%
        \draw[dotted, thick] (0-\del,\spacing*\x)--(\spacing*\rows+\del,\spacing*\x);
        \foreach \y in {0,...,\rows} {
            \fill (\x*\spacing, \y*\spacing) circle (2.5pt);
        }
    }
    \draw[->, ultra thick, green] (\spacing, \spacing) -- (\spacing, 2*\spacing);
    \draw[->, ultra thick, red] (\spacing, \spacing) -- (2*\spacing, \spacing);
    \draw[->, ultra thick, blue] (\spacing, \spacing) -- (0, \spacing);
    \draw[->, ultra thick, magenta] (\spacing, \spacing) -- (\spacing, 0);
    \node at (-1.8*\del,\spacing-\del) {$(i-1,j)$};
    \node at (\spacing+\del,\spacing-\del) {$(i,j)$};
    \node at (2*\spacing-1.5*\del,\spacing-\del) {$(i+1,j)$};
    \node at (\spacing+1.8*\del,\del/2) {$(i,j-1)$};
    \node at (\spacing-\del,2*\spacing+\del) {$(i,j+1)$};
    \node at (1.6*\spacing,\spacing+\del*0.8) {$\textcolor{red}{\cj_x^{i,j}+\delta m/\delta t}$};
    \node at (\spacing/2,\spacing+\del) {$\textcolor{blue}{\cj_x^{i-1,j}-\delta m/\delta t}$};
    \node[rotate=90] at (3.5,4.8) {$\textcolor{green}{\cj_x^{i,j}+\delta m/\delta t}$};
    \node[rotate=90] at (2.5,1.5) {$\textcolor{magenta}{\cj_x^{i,j-1}-\delta m/\delta t}$};
    \end{tikzpicture}
    \caption{The schematic diagram illustrates a mass-update process and the associated instantaneous bond currents across $4$ different bonds, resulting from the loss of mass $m_{i,j} - 4\delta m$ at site $(i,j)$ and the gain of mass $\delta m$ at each of the four neighboring sites in the time interval $(t, t+\delta t)$. During the mass-transfer event, the corresponding instantaneous bond currents are updated as following: $\mathcal{J}_x^{i,j} \rightarrow \mathcal{J}_x^{i,j} + \delta m/\delta t$ (rightward, red arrow), $\mathcal{J}_x^{i-1,j} \rightarrow \mathcal{J}_x^{i-1,j} - \delta m/\delta t$ (leftward, blue arrow), $\mathcal{J}_y^{i,j} \rightarrow \mathcal{J}_y^{i,j} + \delta m/\delta t$ (upward, green arrow), and $\mathcal{J}_y^{i,j-1} \rightarrow \mathcal{J}_y^{i,j-1} - \delta m/\delta t$ (downward, magenta arrow). }
    \label{fig:curr_def}
\end{figure}

In two dimensions, the current can flow along $x$ or $y$ axis (or, along both the axes during an infinitesimal time interval as in this particular case MCM-CoMC I considered in this section). Consequently, the current correlations have two distinct components: the correlation function for currents along the same direction, $C_{r,s}^{\cq_x\cq_x}(t,t^\prime)$ [or, $C_{r,s}^{\cq_y\cq_y}(t,t^\prime)$] and the ``cross''-correlation function, $C_{r,s}^{\cq_x\cq_y}(t,t^\prime)$ or $C_{r,s}^{\cq_y\cq_x}(t,t^\prime)$. For convenience, we use the dynamical correlation for two observables $A_x$ and $B_y$ as $C^{A_xB_y}_{r,s}(t,t^\prime)= \langle A_x^{i,j}(t)B_y^{i+r,j+s}(t^\prime) \rangle_c$ in the rest of the paper.

\subsection{Calculation of current correlation functions}\label{sec:cal_I2d}

\begin{figure*}
    \centering
    \includegraphics[width=0.33\linewidth]{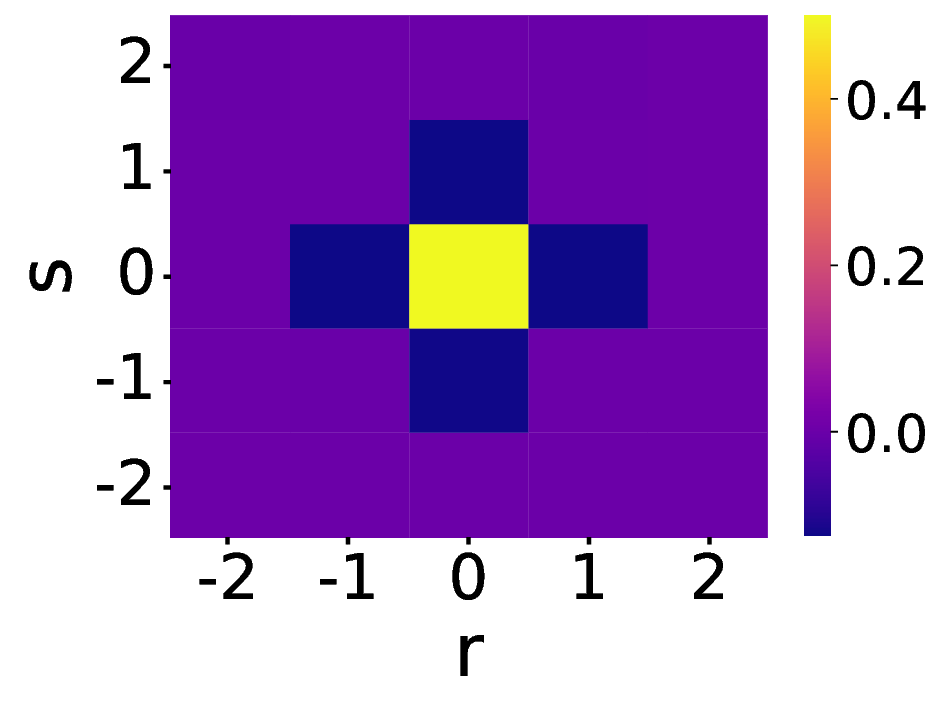}
    \put(-95,128){\textbf{ (a)}}
    \includegraphics[width=0.33\linewidth]{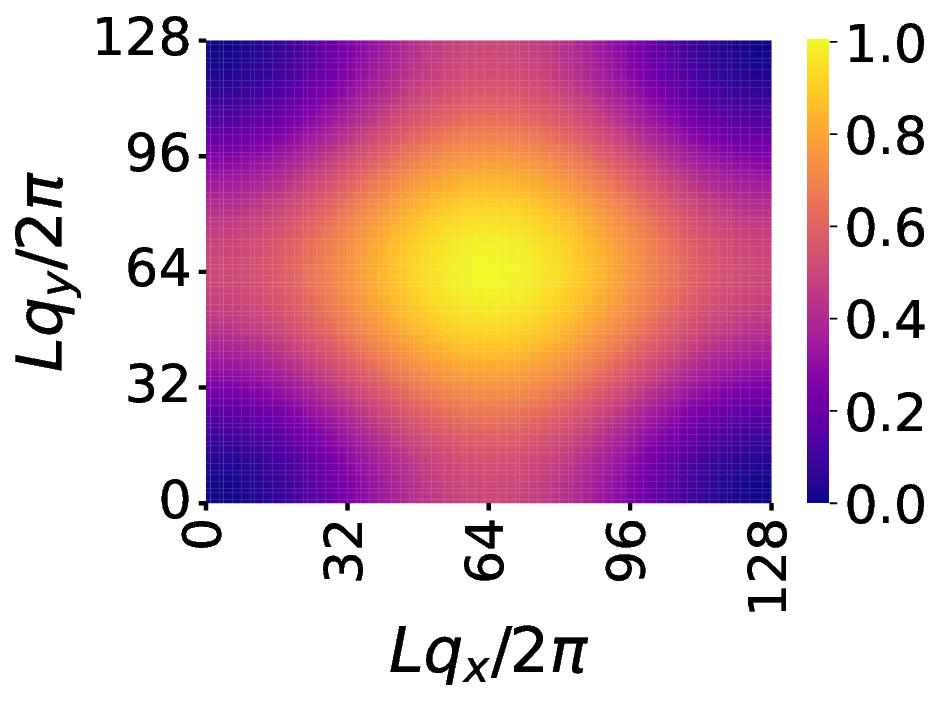}
    \put(-90,125){\textbf{(b)}}
    \includegraphics[width=0.33\linewidth]{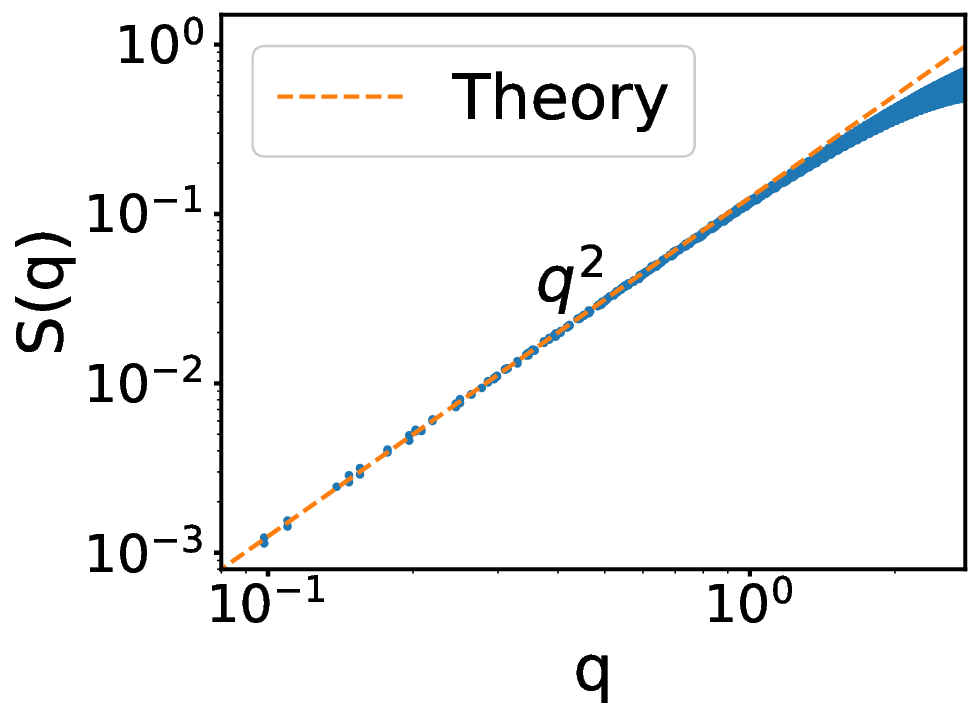}
    \put(-80,125){\textbf{(c)}}
    \caption{ {\it MCM-CoMC I in two dimensions.} Panel (a): Steady-state density correlation $C_{r,s}^{mm}$ is shown in a 2D color heat plot. In this panel, we have taken the periodic system size of area  $100 \times 100$. Panel (b): The heat-map of structure factor $S({\bf q})$ is plotted in the two-dimensional (scaled) ${\bf q}$-plane. Panel
  (c): The structure factor $S(q)$ is shown as a function of the magnitude $q=\sqrt{q_x^2+q_y^2}$. For small-$q$ values, $S(q)$ scales as $q^2$ [see Eq. \eqref{eq:cmmqq}]. In panels (b) and (c), we take a periodic system of area $128 \times 128$. Global density $\rho=1.0$ for all the panels. }
    \label{fig:crsq_MCM_COM_I_2d}
\end{figure*}

The unequal-time $(t>t')$ correlation function for bond currents in the same direction (i.e., the correlation between the current in the $x$ direction at initial time with that at later time), $C_{r,s}^{\cq_x\cq_x}(t,t^\prime)$, which in the Fourier domain ${\bf q} \equiv (q_x,q_y)$ satisfies the following time-evolution equation 
\begin{equation}
\label{eq:cq_xq_x_fr_2dmcmI}
   \frac{d}{dt}\tilde{C}_{q_x,q_y}^{\cq_x\cq_x}(t,t^\prime) = D (1-e^{iq_x})\tilde{C}_{q_x,q_y}^{m\cq_x}(t,t^\prime),
\end{equation}
where $D=\mu_1/4$ is the bulk-diffusion coefficient. 
Here, we denote the Fourier transform of the two dynamical observables $A$ and $B$, in two dimensions, as $$\tilde{C}_{q_x,q_y}^{AB}=\sum_{r,s}C_{r,s}^{AB} e^{iq_x r} e^{iq_y s},$$ and accordingly the inverse Fourier transform is given by $$C_{r,s}^{AB}=\frac{1}{L^2} \sum_{q_x,q_y}\tilde{C}_{q_x,q_y}^{AB} e^{-iq_x r} e^{-iq_y s}.$$ 
Now, to solve Eq. \eqref{eq:cq_xq_x_fr_2dmcmI}, we need to calculate the unequal-time mass-current correlation function in the Fourier domain, $\tilde{C}_{q_x,q_y}^{m\cq_x}(t,t^\prime)$, which satisfies the following equation,
\begin{align}
\label{eq:cmq_x_fr_2dmcmI}
    \frac{d}{dt}\tilde{C}_{q_x,q_y}^{m\cq_x}(t,t^\prime)=-D\omega(q_x,q_y)\tilde{C}_{q_x,q_y}^{m\cq_x}(t,t^\prime),
\end{align}
with
\begin{align}
    \omega(q_x,q_y)=\lambda(q_x)+\lambda(q_y)
\end{align}
and $\lambda(q_\alpha)=2[1-\cos(q_\alpha)]$ with $\alpha \in [x,y]$ . The above-mentioned Eqs. \eqref{eq:cq_xq_x_fr_2dmcmI} and \eqref{eq:cmq_x_fr_2dmcmI} are two first-order ordinary differential equations,  which can be solved exactly by knowing the following equal-time correlations: $\tilde{C}_{q_x,q_y}^{\cq_x\cq_x}(t,t)$ and $\tilde{C}_{q_x,q_y}^{m\cq_x}(t,t)$ as initial conditions. Further, from the microscopic update rules, we calculate the equal-time current-current correlation in Fourier space-domain, $\tilde{C}_{q_x,q_y}^{\cq_x\cq_x}(t,t)$, which satisfies the following equation,
\begin{align}\label{eq:dcqqtt_q_mcm_I2d}
    \frac{d}{dt}\tilde{C}_{q_x,q_y}^{\cq_x\cq_x}(t,t)=2D\tilde{C}_{q_x,q_y}^{m\cq_x}(t,t)(1-e^{iq_x})+\tilde{\Gamma}^{xx}_{q_x,q_y},
\end{align}
where $\tilde{\Gamma}^{xx}_{q_x,q_y}=\mu_1\mu_2\rho^2\lambda(q_x)/8(2\mu_1-\mu_2)$ is the corresponding source term.
Now, using the microscopic update rule, the time evolution of the equal-time mass-current correlation in the Fourier space, i.e., $\tilde{C}_{q_x,q_y}^{m\cq_x}(t,t)$, satisfies the following equation:
\begin{align}\label{eq:dcmqtt_q_mcm_I2d}
    \frac{d}{dt}\tilde{C}_{q_x,q_y}^{m\cq_x}(t,t)=-D\omega(q_x,q_y)\tilde{C}_{q_x,q_y}^{m\cq_x}(t,t)+\tilde{f}^x_{q_x,q_y}.
\end{align}
Here the source term $\tilde{f}^x_{q_x,q_y}= -\mu_1\mu_2\rho^2(1-e^{-iq_x})\omega(q_x,q_y)/16(2\mu_1-\mu_2)$ is evaluated from the knowledge of the steady-state density correlation, which, in Fourier space, can be obtained by solving following equation,
\begin{align}\label{eq:dtcmmqq}
     \frac{d}{dt}\tilde{C}^{mm}_{q_x,q_y}(t,t) = -2D\omega\tilde{C}^{mm}_{q_x,q_y}(t,t) + \omega^2\frac{\mu_2}{16}\langle m^2 \rangle.
\end{align}
Here we have the bulk-diffusion coefficient $D=\mu_1/4$ and the second moment of mass $m_i$ at a site $i$ as $\langle m^2\rangle \equiv  \int  m^2 {\rm Prob.}[m_i = m] dm$.
In the steady-state, we have $d\tilde{C}^{mm}_{q_x,q_y}(t,t)/dt=0$, and we obtain the desired density correlation function as follows:  
\begin{align}\label{eq:cmmqq}
    \tilde{C}_{q_x,q_y}^{mm} = \rho S(q_x,q_y)&= \frac{\mu_2\rho^2}{4(2\mu_1-\mu_2)}\omega(q_x,q_y)\\ \nonumber &\simeq \frac{\mu_2\rho^2}{4(2\mu_1-\mu_2)}(q_x^2+q_y^2) ,
\end{align}
where the second moment of on-site mass can be explicitly calculated as $ \langle m^2\rangle = 2\mu_1\rho^2/(2\mu_1-\mu_2)$.
Notably, the Fourier transform of the density correlation in Eq. \eqref{eq:cmmqq} is related to the structure factor $S(q)$ up to a constant factor. This implies that, in the small-$q$ limit, where $\lambda(q_x) \sim q_x^2$ as $q_x \to 0$, the structure factor $S(q)$ scales as $q^2$, with $q = \sqrt{q_x^2 + q_y^2}$. 
Indeed, the vanishing of the static structure factor $S(q)$ at small wave number is a signature of extreme (``class-I") hyperuniformity of spatial density fluctuations \cite{Torquato2018Jun}.  
In Fig. \ref{fig:crsq_MCM_COM_I_2d}, we provide a heat-map of the steady-state density correlation and structure factor for the two-dimensional mass chipping model. Panel (a) presents a two-dimensional color heat map of the steady-state density correlation function $C_{r,s}^{mm}$ for an MCM on a periodic square lattice with an area of $100 \times 100$. In Panel (b), we show the heat map of the structure factor, $S(q_x, q_y)$, in the scaled two-dimensional plane $(Lq_x/2\pi, Lq_y/2\pi)$. Panel (c) depicts the structure factor $S(q)$ as a function of the magnitude $q = \sqrt{q_x^2 + q_y^2}$ of the wave number vector ${\bf q} \equiv \{ q_x, q_y\}$. For small $q$ values, $S(q)$ exhibits a scaling behavior proportional to $q^2$, as described by Eq. \eqref{eq:cmmqq}. In all panels, the structure factors calculated in the simulations are for an MCM on a periodic lattice with an area of $128 \times 128$ and global density $\rho = 1.0$.

We provide below the results for dynamic correlations in the following two cases. In the first case, we calculate the correlation function for currents along the same direction (say, both along $x$ direction). In the second case, we do the same for currents along two perpendicular directions (say, along $x$ and $y$ directions).

\subsection{Case I: Correlation between currents along the same direction.}

Finally, we obtain a closed set of Eqs. 
\eqref{eq:cq_xq_x_fr_2dmcmI}, \eqref{eq:cmq_x_fr_2dmcmI}, \eqref{eq:dcqqtt_q_mcm_I2d}, and \eqref{eq:dcmqtt_q_mcm_I2d} from which, after some algebraic manipulation (e.g., few integrations followed by inverse Fourier transform, which is defined as $C_{r,s}^{AB}=1/L^2\sum_{q_x,q_y}\tilde{C}^{AB}_{q_xq_y}$), we are able to calculate the time-integrated bond-current fluctuation of unequal-time $C_{r,s}^{\cq_x\cq_x}(t,t^\prime)$. However, by setting  $r=s=0$ and $t=t^\prime=T$, the second cumulant, or the variance, of the time-integrated current $\cq_x(T)$  across a particular bond in $x$ direction and in time interval $T$ can be obtained from $C_{0,0}^{\cq_x\cq_x}(T, T) = \langle [\cq_x(T)]^2\rangle_c$, and it can be expressed explicitly as
\begin{widetext}
\begin{align}
\label{eq:c00tt_com1_2d_sum}
   \langle [\cq_x(T)]^2 \rangle_c = C_{0,0}^{\cq_x\cq_x}(T,T) = T\frac{\mu_2 \langle m^2\rangle}{8}+\frac{2D}{L^2}\sum_{q_x,q_y}\left[\frac{T}{D\omega}-\frac{(1-e^{-D\omega T})}{(D\omega)^2}\right]\lambda_{q_x}\left[D\tilde{C}^{mm}_{q_x,q_y}-\frac{\mu_2}{16}\langle m^2 \rangle {\omega}\right],
\end{align}
\end{widetext}
where the summation is over $(q_x,q_y) \ne (0,0) $.
The above equation in the long-time regime can be written in the following asymptotic form,
\begin{align}
\label{2D-A}
  \langle [\cq_x(T)]^2\rangle_c = \langle [\cq_x(T)]^2\rangle  \simeq A_1T +A_2 + \frac{A_3}{T}.
\end{align}
After some algebraic manipulation, we obtain from Eq. \eqref{eq:c00tt_com1_2d_sum} the first coefficient,  
\begin{align}
\label{A1-zero-2D}
    A_1(\rho) = \frac{\mu_2 \langle m^2\rangle}{32}\left[4- \frac{2}{L^2}\sum_{q_x, q_y}\lambda_{q_x}\right]=0,
\end{align}
which, quite interestingly, vanishes as in the case of one-dimensional models discussed in the previous sections; here, we have used the identity $\sum_{q_x}\lambda_{q_x} = 2L$. In other words, the coefficient of the leading-order term of Eq. \eqref{2D-A} having linear ($\sim T$) growth is zero; however, as we show later, this particular behavior may not in general be observed in two (and higher) dimensions and depends on the microscopic details of the models under consideration. Now, in the thermodynamic limit $L \to \infty$ and keeping density fixed, we can express Eq. \eqref{eq:c00tt_com1_2d_sum} as the following integral,
\begin{align}\label{eq:c00tt_com1_2d}
    \langle [\cq_x(T)]^2\rangle_c &= \frac{\mu_2\rho^2}{2(2\mu_1-\mu_2)}\frac{1}{(2\pi)^2}\\ \nonumber &\times\int\limits_0^{2\pi} \int\limits_0^{2\pi} dq_x dq_y \frac{\lambda (q_x)\left( 1-e^{-D\omega T}\right)}{\lambda (q_x)+\lambda (q_y)},
\end{align}
For large time $T$, we have the asymptotic expression of Eq. \eqref{eq:c00tt_com1_2d} as
\begin{align}\label{eq:Cqxqx_comI_asym}
    C_{0,0}^{\cq_x\cq_x}(T,T)\simeq A_2(\rho)\left[1-\frac{1}{4\pi D}  T^{-1}\right],
\end{align}
where we have used the approximation $\lambda(q_\alpha)\approx q_\alpha^2$ and obtain 
\begin{align}\label{eq:qsat_2D_comI}
 A_2(\rho) =   \lim_{T\to\infty}C_{0,0}^{\cq_x\cq_x}(T, T)= \frac{\mu_2\rho^2}{4(2\mu_1-\mu_2)}.
\end{align}
In other words, at long times, the variance of time-integrated bond current, quite strikingly, saturates to a constant value.
Next, we calculate the two-point unequal-time correlation function for bond current (correlation between the currents at unequal times, but both along the same, say $x$, direction), which can be obtained as
\begin{widetext}
\begin{align}\label{eq:cqxx_comI_th}
    C_{0,0}^{\cj_x\cj_x}(t,0) =\left[\frac{d}{dt}\frac{d}{dt^\prime}C_{0,0}^{\cq_x\cq_x}(t,t^\prime)\right]_{t^\prime=0, t\ge0}= \Gamma_{0, 0}^{xx}\delta(t)-\frac{\mu_1^2\mu_2\rho^2}{64(2\mu_1-\mu_2)}\frac{1}{L^2}\sum_{q_x,q_y}e^{-D\omega t}\lambda_{q_x} \omega.
\end{align}
By taking the infinite-volume limit $L\to \infty$ with $t > 0$ finite, the above expression can be written in an integral form,
\begin{align}\label{eq:j0jt_xx_com_2D_int}
     C_{0,0}^{\cj_x\cj_x}(t,0)=- \frac{\mu_1^2\mu_2\rho^2}{64(2\mu_1-\mu_2)}\frac{1}{(2\pi)^2}\int\limits_0^{2\pi} \int\limits_0^{2\pi} dq_x dq_y e^{-D\omega t}\lambda(q_x) \omega(q_x, q_y).
\end{align}\end{widetext}
With the use of the approximation $\lambda(q_\alpha)\approx q_\alpha^2$ in the above equation, we get the asymptotic behavior of instantaneous current correlation function at large times,
\begin{align}\label{eq:cqxx_comI}
    C_{0,0}^{\cj_x\cj_x}(t,0) \simeq - \frac{\mu_2\rho^2}{4\pi\mu_1(2\mu_1-\mu_2)}t^{-3}.
\end{align}
Clearly, as in Eq. \eqref{eq:cqxx_comI}, the decay of the dynamic (two-point) correlation function for bond current in {\it two} dimensions differs from that of the CoM-conserving models in {\it one} dimension as shown in Eq. \eqref{eq:cjjtasym_com1d} for MCMs and in Ref. \cite{Mukherjee2024Aug} for the Oslo model.

\subsection{Case II: Correlation between currents along two perpendicular directions.}

\begin{figure*}
    \centering
    \includegraphics[width=0.33\linewidth]{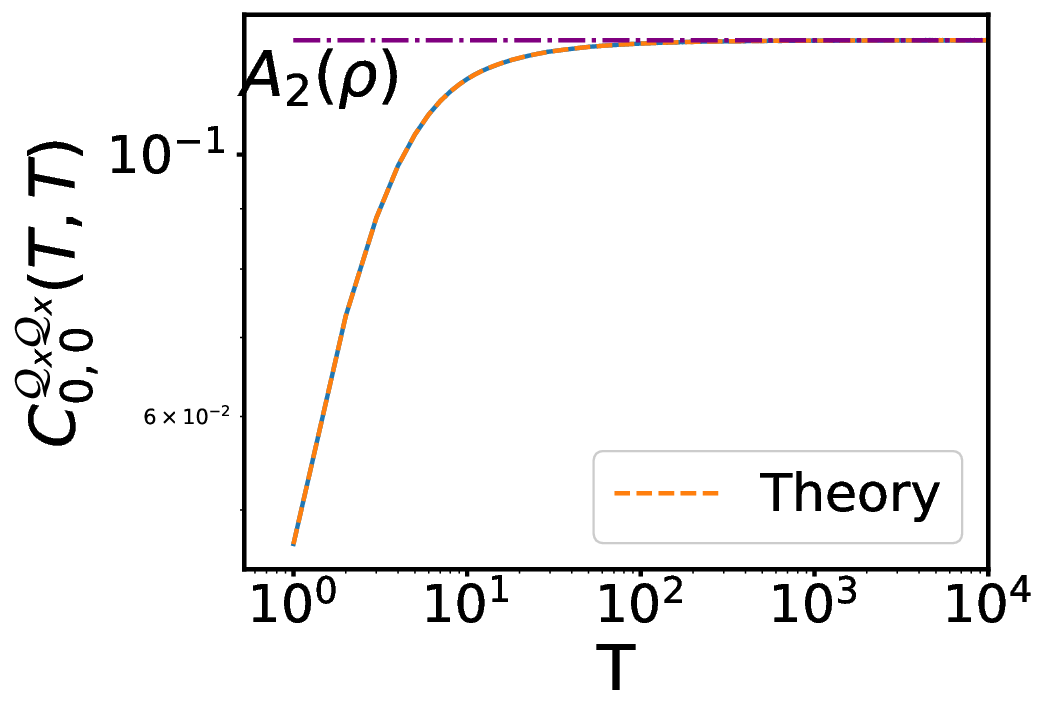}
    \put(-25,72){\textbf{(a)}}
    \includegraphics[width=0.33\linewidth]{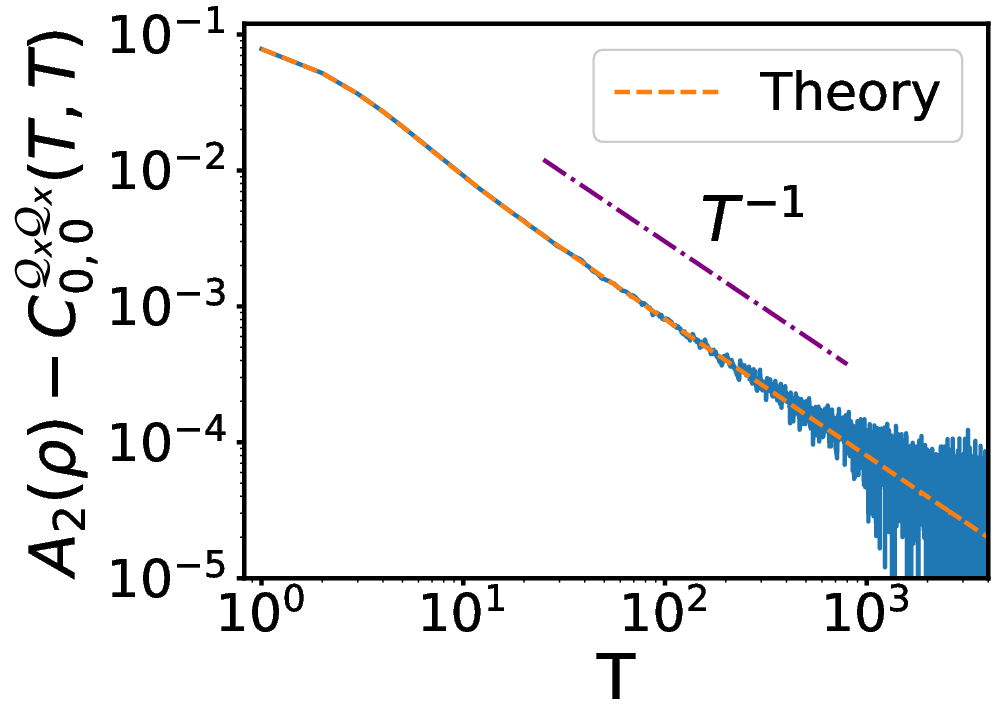}
    \put(-80,60){\textbf{(b)}}
    \includegraphics[width=0.33\linewidth]{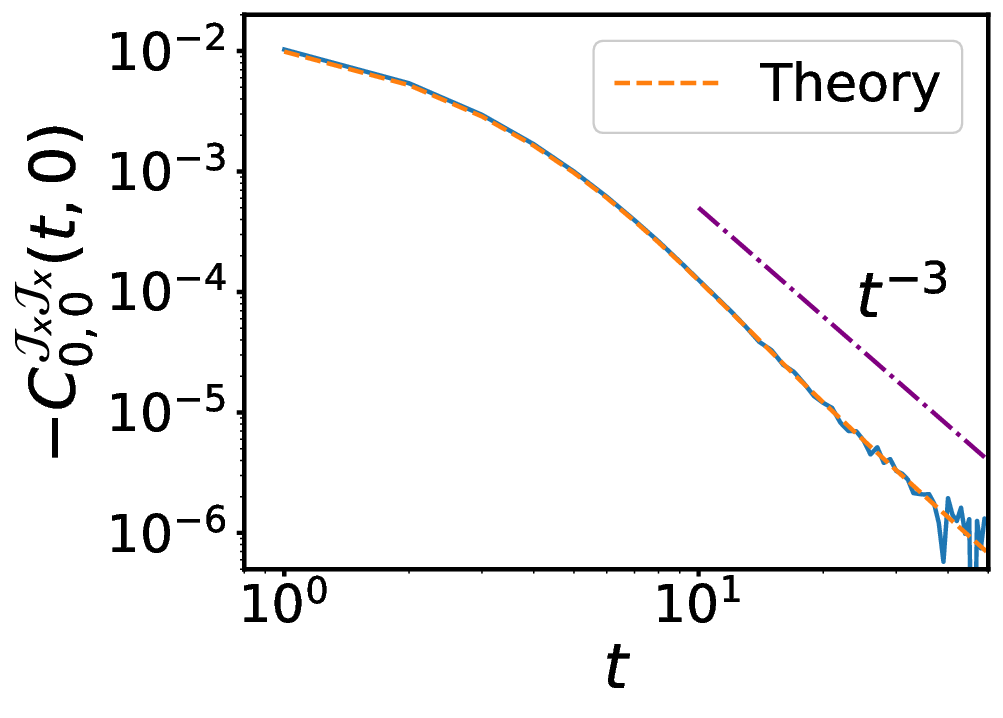}
    \put(-90,70){\textbf{(c)}}\\
    \includegraphics[width=0.33\linewidth]{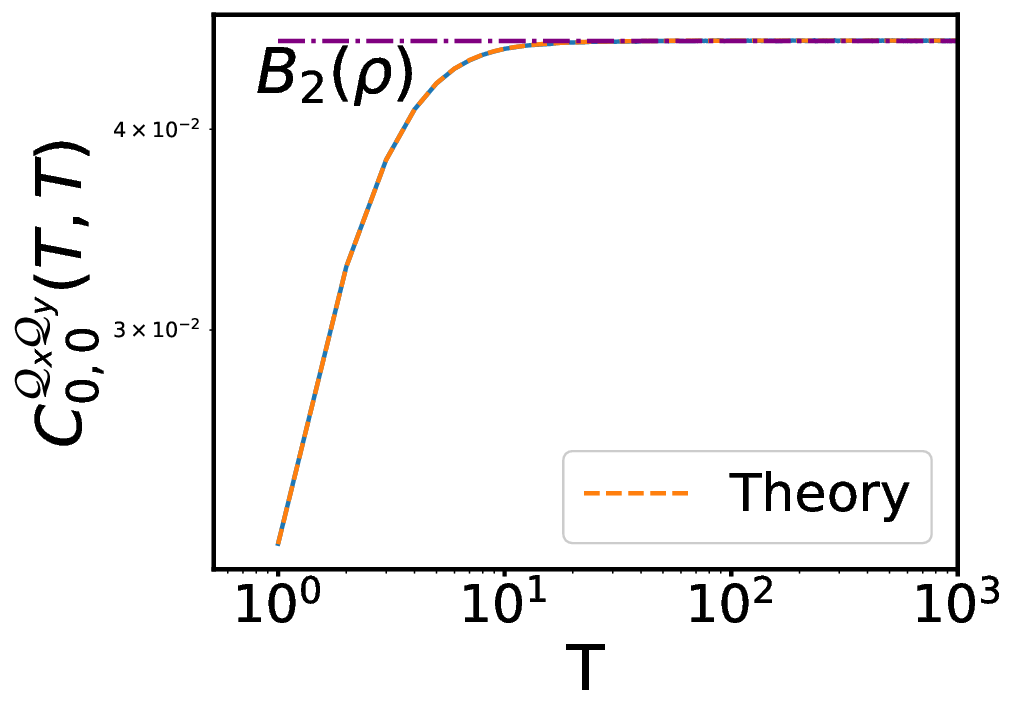}
    \put(-25,75){\textbf{(d)}}
    \includegraphics[width=0.33\linewidth]{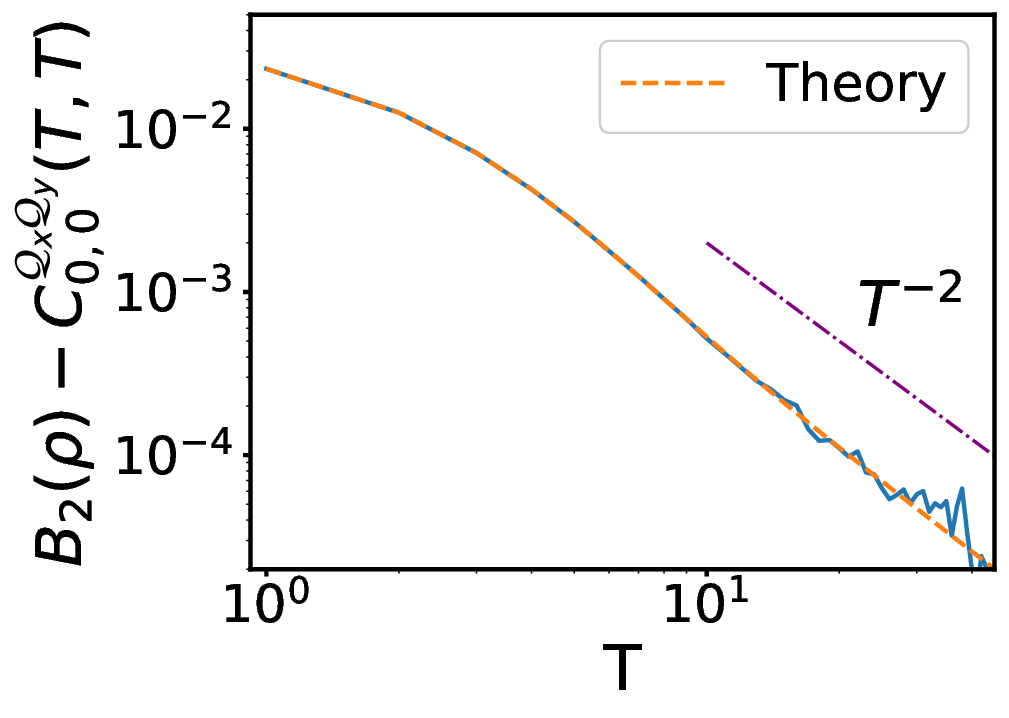}
   \put(-80,65){\textbf{(e)}} 
   \includegraphics[width=0.33\linewidth]{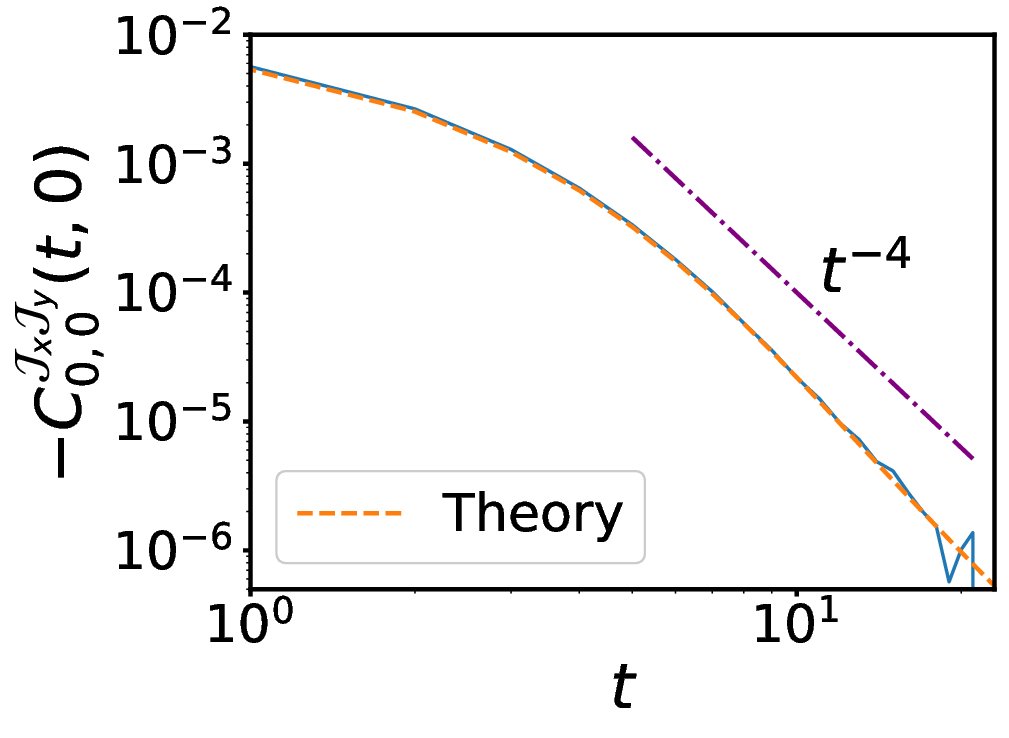}
    \put(-90,70){\textbf{(f)}} 
    
    \caption{  {\it MCM-CoMC I in two dimensions:} Panel (a): Equal-time bond-current fluctuations (in the same direction) $C_{0, 0}^{\cq_x\cq_x}(T,T)$ are plotted as a function of time $T$.  Panel (b): The relative bond current fluctuation  $A_2(\rho)- C_{0, 0}^{\cq_x\cq_x}(T,T)$, is plotted as a function of time $T$. The decay follows a $T^{-1}$ [see Eq. \eqref{eq:Cqxqx_comI_asym} ] behavior (purple dot-dashed). Panel (c): The instantaneous current correlation in the same direction, $C_{0,0}^{\cj_x\cj_x}(t,0)$, is plotted as a function of time $t$. The cross-correlation of the bond current, $C_{0, 0}^{\cq_x\cq_y}(T,T)$ (panel d), and the asymptotic behavior of $A_2(\rho)- C_{0, 0}^{\cq_x\cq_y}(T,T)$ (panel e) are plotted as a function of time $T$. Panel (f): The cross current correlation, $C_{0,0}^{\cj_x\cj_y}(t,0)$, is plotted as a function of time $t$. The dashed lines (orange) in panels (a), (c), (d), and (f) are obtained from theory as in Eqs. \eqref{eq:c00tt_com1_2d}, \eqref{eq:j0jt_xx_com_2D_int}, \eqref{eq:cqxytt_comI_int} and \eqref{eq:cjxjy_comI_int}, respectively. In all panels, simulations (blue line) have been performed for the system of area $100\times 100$ and global density $\rho=1.0$.} 
    \label{fig:all_2d_comI}
\end{figure*}

The unequal-time $(t>t')$ current correlation between currents along two perpendicular directions $x$ and $y$ can be written as
\begin{equation}
\label{eq:cq_xq_xy_fr_2dmcmI}
   \frac{d}{dt}\tilde{C}_{q_x,q_y}^{\cq_x\cq_y}(t,t^\prime) = D (1-e^{iq_x})\tilde{C}_{q_x,q_y}^{m\cq_y}(t,t^\prime),
\end{equation}
where unequal-time mass current correlation $\tilde{C}_{q_x,q_y}^{m\cq_y}(t,t^\prime)$ satisfy following equation
\begin{align}
\label{eq:cmq_y_fr_2dmcmI}
    \frac{d}{dt}\tilde{C}_{q_x,q_y}^{m\cq_y}(t,t^\prime) = - D \omega(q_x,q_y)\tilde{C}_{q_x,q_y}^{m\cq_y}(t,t^\prime),
\end{align}
Now to solve Eq. \eqref{eq:cq_xq_xy_fr_2dmcmI}, we require equal-time  correlation function for two cumulative currents $\cq_x$ and $\cq_y$ in the perpendicular directions $x$ and $y$, respectively, 
\begin{align}\label{eq:qqtt_xy_I}
    \frac{d}{dt} \tilde{C}^{\cq_x \cq_y}_{q_x,q_y}(t,t)&=D(1-e^{iq_x})\tilde{C}^{m\cq_y}_{q_x,q_y}(t,t)\\ \nonumber &+D(1-e^{iq_y})\tilde{C}^{m\cq_x}_{q_x,q_y}(t,t)+\tilde{\Gamma}^{xy}_{q_x,q_y}(\rho),
\end{align}
where ${\tilde \Gamma}^{x y}_{q_x,q_y}(\rho) = \mu_1\mu_2\rho^2(1-e^{iq_x})(1-e^{-iq_y})/8(2\mu_1-\mu_2)$.
Furthermore, following along the lines of the calculations given in section \ref{sec:cal_I2d} and explicitly solving the above set of time-evolution equations, 
we calculate equal-time ``cross''-correlation function (i.e., correlation between currents in $x$ and $y$ directions, measured up to time $T$) for the time-integrated bond current (by setting $r=s=0$),
\begin{align}\label{eq:cqxytt_comI}
    &C_{0,0}^{\cq_x\cq_y}(T, T)\\ \nonumber &= \frac{\mu_2\rho^2}{4(2\mu_1-\mu_2)} \frac{1}{L^2}\sum_{(q_x,q_y) \ne (0,0)}\frac{(1-e^{-D\omega T})}{\omega}\Lambda(q_x,q_y)
    \\ \nonumber
    &\simeq B_1 T  + B_2 + B_3 T^{-2},
\end{align}
where we denote $$\Lambda(q_x,q_y)=[\lambda(q_x)+\lambda(q_y)-\lambda(q_x-q_y)],$$
and $B_1$, $B_2$, and $B_3$ are density-dependent constant coefficients, describing the asymptotic growth law for cross-correlation function. 
Note that, in this particular case of two-dimensional MCM-CoMC I,  the coefficient $B_1$ of the first term corresponding to the linear-time growth is identically zero, and we find 
\begin{align}
\label{B1-zero}
    B_1 &= \frac{\mu_2 \langle m^2\rangle}{32} \left[2 - \frac{1}{L^2}\sum_{(q_x,q_y) \ne (0,0)} \Lambda(q_x,q_y) \right]=0;
\end{align}
this is, however, not the case for the other two-dimensional case of MCM-CoMC II discussed in the next section and depends on microscopic details.
In the thermodynamic limit ($L\to \infty$ and global density $\rho$ being fixed), the summation in the above equation can be written as an integral,
\begin{align}
\label{eq:cqxytt_comI_int}
    &C_{0,0}^{\cq_x\cq_y}(T, T)\\ \nonumber &= \frac{\mu_2\rho^2}{4(2\mu_1-\mu_2)} \frac{1}{(2\pi)^2} \int\limits_0^{2\pi} \int\limits_0^{2\pi} dq_x dq_y \frac{(1-e^{-D\omega T})}{\omega}\Lambda(q_x, q_y).
\end{align}
The form of the second factor $\Lambda(q_x,q_y) = \lambda(q_x)+\lambda(q_y) - \lambda(q_x-q_y)$ in the right hand side (inside the integral) of the above equation is generic in any dimension $d$, where one considers correlations between currents along any two perpendicular directions (here, $x$ and $y$); see the sketch of the calculation details given in \ref{app:ddim_cal}. 
In the above expression, we can write the factor explicitly as $\Lambda(q_x,q_y) = 2(1-\cos(q_x))+2(1-\cos(q_y))-2+2\cos (q_x)\cos(q_y)+2\sin(q_x)\sin(q_y)$. Now the integration over the last term $2 \sin(q_x) \sin(q_y)$ in the factor $\Lambda$ vanishes. Therefore, in the limit of small wave numbers, the remaining terms, in the leading order, can be approximated as $\lambda(q_x) + \lambda(q_y) - \lambda(q_x - q_y)\approx q_x^2 q_y^2/2$, leading to the following asymptotic form of the ``cross-correlation'' function for cumulative bond currents, 
\begin{align}
    C_{0,0}^{\cq_x\cq_y}(T, T) \simeq B_2(\rho) - \frac{2\mu_2\rho^2}{16\mu_1^2(2\mu_1-\mu_2)\pi}T^{-2},
\end{align}
which, for large times, saturates to a density-dependent constant,
\begin{align}
    B_2(\rho) = \frac{\mu_2\rho^2}{4(2\mu_1-\mu_2)}\frac{1}{(2\pi)^2}\int\limits_0^{2\pi} \int\limits_0^{2\pi} dq_x dq_y \frac{\Lambda(q_x,q_y) }{\omega}.
\end{align} 
Notably, in the above equation, the dynamic ``cross''-correlation function  $C_{0,0}^{\cq_x\cq_y}(T, T)$ for time-integrated bond currents along two orthogonal ($x$ and $y$) directions decays much faster, i.e., as a $T^{-2}$ power law [see panel (e) of Fig. \ref{fig:all_2d_comI}], as compared to that along the same direction, which decays as $T^{-1}$ [see panel (b) of Fig. \ref{fig:all_2d_comI}] obtained in Eq. \eqref{eq:Cqxqx_comI_asym}.
Now, we calculate the dynamical cross-correlation function for the instantaneous bond current as 
\begin{widetext}
\begin{align}\label{eq:cjxjy_comI_th}
    C_{0,0}^{\cj_x\cj_y}(t,0) = \Gamma_{0,0}^{xy}\delta(t)- \frac{\mu_1^2\mu_2\rho^2}{128(2\mu_1-\mu_2)}\frac{1}{L^2}\sum_{q_x,q_y} e^{-D \omega(q_x, q_y) t} \omega(q_x, q_y) \Lambda(q_x, q_y),
\end{align}
where $\Gamma_{0,0}^{xy}=\mu_2\langle m^2\rangle/16$.
Now, in the thermodynamic limit $L \to \infty$, we obtain, for time $t>0$,
\begin{align}\label{eq:cjxjy_comI_int}
     C_{0,0}^{\cj_x\cj_y}(t,0) = -\frac{\mu_1^2\mu_2\rho^2}{128(2\mu_1-\mu_2)}\frac{1}{(2\pi)^2}\int\limits_0^{2\pi} \int\limits_0^{2\pi} dq_x dq_y e^{-D \omega(q_x, q_y) t} \omega (q_x, q_y) \Lambda(q_x, q_y)\simeq -\frac{3\mu_2\rho^2}{16\mu_1^2(2\mu_1-\mu_2)\pi}t^{-4}.
\end{align} 
\end{widetext}

In panel (a) of Fig. \ref{fig:all_2d_comI}, we plot the variance  $\langle [\cq^x(T)]^2\rangle_c \equiv C_{0, 0}^{\cq_x \cq_x}(T,T)$ of time-integrated bond current along $x$ direction in time interval $T$ as a function of time $T$.  In panel (b) of Fig. \ref{fig:all_2d_comI}, the relative bond-current fluctuation  $[A_2(\rho) - C_{0, 0}^{\cq_x\cq_x}(T,T)]$, is plotted as a function of time $T$. The decay follows a $T^{-1}$ power-law behavior as in Eq. \eqref{eq:Cqxqx_comI_asym} (purple dot-dashed). In panel (c) of the same figure, the correlation function  $C_{0,0}^{\cj_x\cj_x}(t,0)$ for bond currents in the same ($x$) direction is plotted as a function of time $t$. The corresponding cross-correlation function $C_{0, 0}^{\cq_x\cq_y}(T,T)$ for time-integrated bond currents along the orthogonal ($x$ and $y$) directions and the asymptotic behavior of $A_2(\rho)- C_{0, 0}^{\cq_x\cq_y}(T,T)$ are plotted as a function of time $T$ in panels (d) and (e), respectively. In panel (f), the cross correlation function $C_{0,0}^{\cj_x\cj_y}(t,0)$ for bond current is plotted as a function of time $t$. The dashed-lines (orange) in panels (a), (c), (d), and (f) are obtained from theory as in Eqs. \eqref{eq:c00tt_com1_2d}, \eqref{eq:j0jt_xx_com_2D_int}, \eqref{eq:cqxytt_comI_int} and \eqref{eq:cjxjy_comI_int}, respectively. In all the panels, simulations (blue line) are for periodic system of area $100\times 100$ and global density $\rho=1.0$.

\section{Two-dimensional MCM-CoMC II}\label{Sec:2D_MCM_COM_II}

In this section, we consider the following model in two-dimensional square lattices on a periodic domain. Mass at a site $(i,j)$ is fragmented with unit rate, where the site retains a random fraction  $\xi m_{i,j}$ of mass. The remaining mass is chipped off and divided into two equal parts,  each of which, having $\tilde{\xi} m_{i,j}/2$ amount of mass, is transferred to each of the two nearest neighbor sites in the horizontal or vertical direction, i.e., either to $(i+1,j)$ and $(i-1,j)$ (horizontal direction) or to $(i,j+1)$ and $(i,j-1)$ (vertical direction), with equal probability.
From the time evolution of the density profile, we were able to calculate the bulk-diffusivity $D=\mu_1/4$, which is the same as the value obtained in the previous model discussed in Section \ref{Sec:2D_MCM_COM_I}.

\subsection{Correlation between currents along the same directions.}

\begin{figure*}
    \centering
     \includegraphics[width=0.33\linewidth]{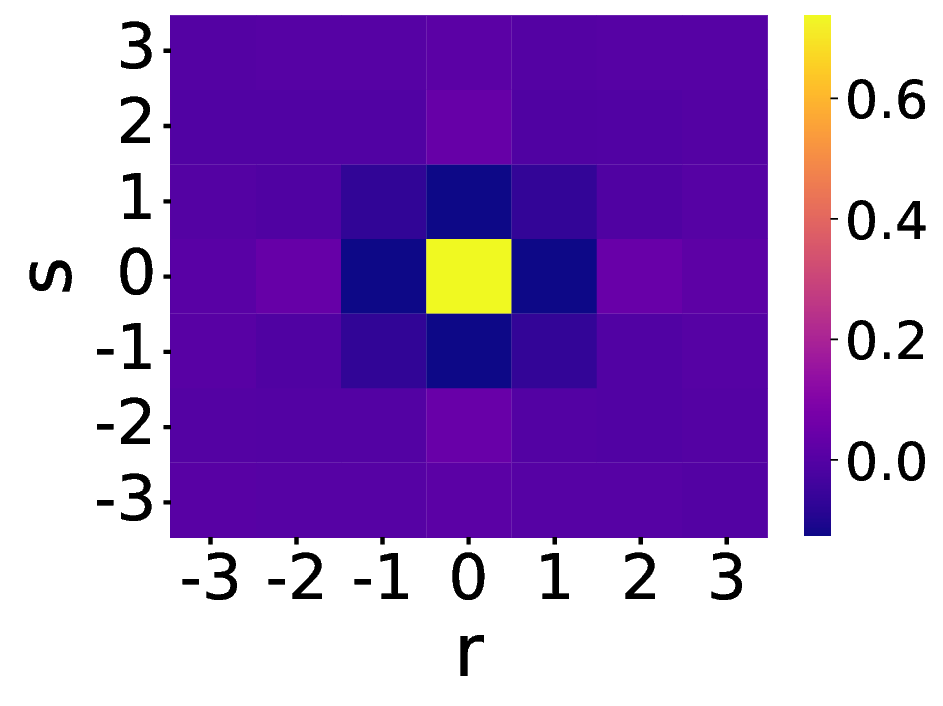}
    \put(-90,128){ \textbf{(a)}}
    \includegraphics[width=0.33\linewidth]{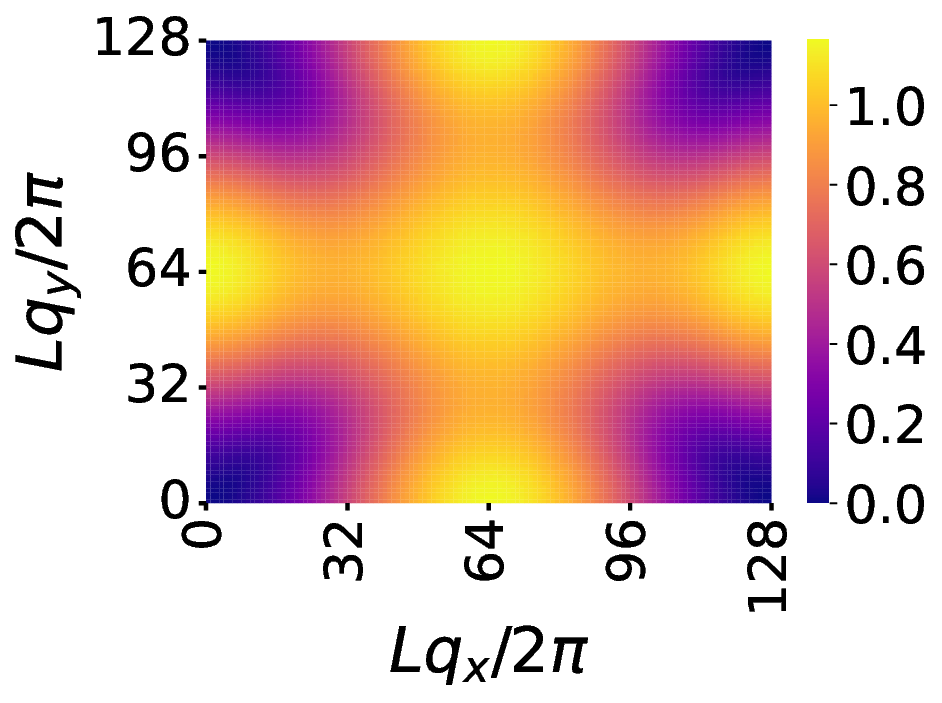}
    \put(-95,125){\textbf{ (b)}}
    \includegraphics[width=0.33\linewidth]{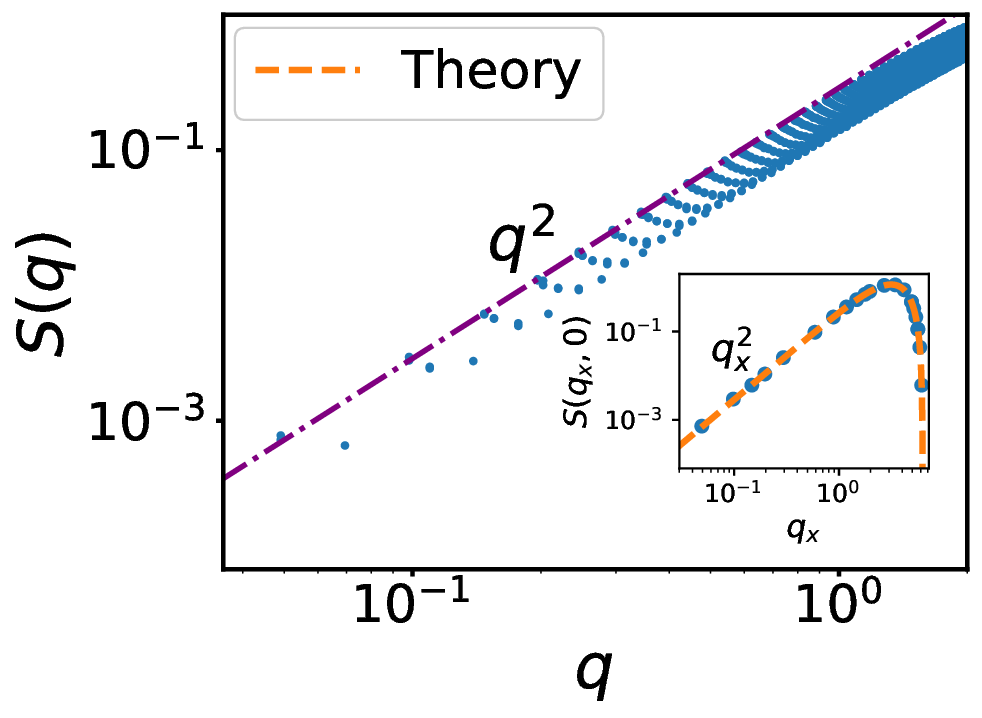}
    \put(-80,125){\textbf{ (c)}}
    \caption{Panel (a): The steady-state density correlation $C_{r,s}^{mm}$ is depicted in a 2D color heat plot. In this panel, we have taken a periodic square lattice of area $100\times 100$. Panel (b): The heat-map of steady-state structure factor $S(\textbf{q})$ is plotted in two-dimensional (scaled) $\textbf{q}$-plane. Panel (c): The structure factor $S(q)$ is plotted as a function of $q$. The guiding line (purple dashed-dotted) represents $q^2$ growth as $q \to 0$. In the inset of panel (c), we have plotted the structure factor $S(q_x,0)$ as a function of $q_x$, setting $q_y=0$ and compared it with theory (orange line) obtained from Eq. \eqref{eq:cqsq_COMII}. In panels (b) and (c), we take a periodic square lattice of area $ L \times L$ with $L=128$ and global density $\rho = 1.0$ for the above-mentioned plots.}
    \label{fig:cor_struc_mcm_com_2d_II}
\end{figure*}

In this section, we calculate the dynamical correlation of current (correlation in the same direction), i.e., $C_{0,0}^{\cq_x\cq_x}$, in terms of the Fourier modes of the density correlation function $\tilde{C}_{q_x, q_y}^{mm}$ as:
\begin{widetext}
\begin{align}\label{eq:cqcq_xx_COMII_th}
    C_{0,0}^{\cq_x\cq_x}(T,T) = T\frac{\mu_2 \langle m^2\rangle}{4} +\frac{2D}{L^2}\sum_{q_x,q_y}\left[\frac{T}{D\omega}-\frac{(1-e^{-D\omega T})}{(D\omega)^2}\right]
\lambda_{q_x}\left[D\tilde{C}^{mm}_{q_x,q_y}-\frac{\mu_2}{8}\langle m^2 \rangle \lambda_{q_x}\right] ,
\end{align}
\end{widetext}
where $\tilde{C}^{mm}_{q_x,q_y}$ is Fourier modes of the steady-state density correlation function and $ C_{0,0}^{mm} \equiv \langle m^2 \rangle > 0$ is the second moment of mass at a single site.
In the long-time regime, the above equation can be written in an asymptotic form,
\begin{align}
 C_{0,0}^{\cq_x\cq_x}(T,T) = \langle [\cq_x(T)]^2 \rangle_c \simeq A_1 T + A_2 + A_3 T^{-1},
\end{align}
where $A_1$, $A_2$ and $A_3$ are constant coefficients.
We note that the coefficient of the first term corresponding to the linear-growth is {\it nonzero} and can be expressed as
\begin{align}
\label{A1-nonzero-2D}
    A_1 =\frac{\mu_2\langle m^2 \rangle}{8}\left[2-\frac{1}{L^2}\sum_{q_x,q_y} \frac{\lambda^2(q_x)}{\omega}\right]\\ \nonumber \approx \frac{\mu_2\langle m^2 \rangle (\pi-2)}{4\pi}\ne 0.
\end{align}
In Eq. \eqref{eq:cqcq_xx_COMII_th}, we have used the Fourier modes of the steady-state density correlation function $\tilde{C}^{mm}_{q_x,q_y}$, which can in principle be obtained from the following equation,
\begin{align}\label{eq:dtcmmqq_COMII}
     \frac{d}{dt}\tilde{C}^{mm}_{q_x,q_y}(t,t) = -2D\omega\tilde{C}^{mm}_{q_x,q_y}(t,t) + (\lambda_{q_x}^2+\lambda_{q_y}^2)\frac{\mu_2}{8}\langle m^2 \rangle.
\end{align}
In the limit $t \to \infty$, the above equation leads to the expression of the density correlation function in the Fourier space, 
\begin{align}\label{eq:cqsq_COMII}
    \tilde{C}^{mm}_{q_x,q_y} = \frac{\mu_2}{4\mu_1}\langle m^2\rangle\frac{\lambda_{q_x}^2+\lambda_{q_y}^2}{\lambda_{q_x}+\lambda_{q_y}}= \rho S(q_x,q_y).
\end{align}
In the subsequent analysis, where we obtain various asymptotic behaviors for bond current fluctuation and the corresponding exponents, we do not actually require to explicitly solve for $\langle m^2 \rangle$, which simply appears as a proportionality constant.  In panel (a) of Fig. \ref{fig:cor_struc_mcm_com_2d_II}, we have shown the heat map of steady-state density correlation $C_{r,s}^{mm}$ for system size $100\times 100$. In panel (b), the heat plot of steady-state structure factor $S(\textbf{q})$ has been plotted in two-dimensional (scaled) ${\bf q}$-plane.  Panel (c): The structure factor $S(q)$ is plotted as a function of $q=\sqrt{q_x^2+q_y^2}$. The guiding line (purple dashed-dotted) represents $q^2$ growth.  In the inset of panel (c), we have plotted the structure factor $S(q_x,0)$ as a function of $q_x$, setting $q_y=0$ and theory line (orange) obtained from Eq. \eqref{eq:cqsq_COMII}. We have measured structure factor for periodic square lattice of area $ 128 \times 128$. Global density has been taken to be $\rho = 1.0$ for all panels.

Furthermore, after some algebraic manipulations,  we can rewrite Eq. \eqref{eq:cqcq_xx_COMII_th} in the following asymptotic form,
\begin{widetext}
\begin{align}\label{eq:cqxx_2D_COMII}
    C_{0, 0}^{\cq_x\cq_x}(T, T) = \langle (\cq_x)^2 \rangle &= A_1T+\frac{\mu_2\langle m^2 \rangle}{(2\pi)^2}\int\limits_0^{2\pi}\int\limits_0^{2\pi} dq_xdq_y(1-e^{-D\omega T})\left[\frac{\lambda(q_x)}{\omega}\right]^2\\ \nonumber
    & \simeq \frac{\mu_2\langle m^2 \rangle (\pi-2)}{4\pi}T + \frac{\mu_2 \langle m^2 \rangle}{2\pi}(\pi-1)-\frac{3\mu_2 \langle m^2 \rangle}{8\mu_1\pi}T^{-1}
    \\ \nonumber
    & \equiv A_1 T + A_2 + A_3 T^{-1},
\end{align}
\end{widetext}
where we use the approximation $\lambda(q_\alpha)\approx q_\alpha^2$.
From the above equation, we find that the variance $C_{0,0}^{\cq_x\cq_x}(T, T)$ of time-integrated bond current grows linearly  $\sim T$ with time; indeed, this particular growth law is qualitatively different from the other variants of one- and two-dimensional MCM-CoMC studied in the previous sections \ref{Sec:1D_MCM_COM_I}, \ref{Sec:1D_MCM_COM_II}, and  \ref{Sec:2D_MCM_COM_I}, where $A_1$ is zero and the variance of time-integrated bond current saturate to a constant value at long times.
We next calculate the correlation function, $C_{0, 0}^{\cj_x\cj_x}(t,0)\equiv C_{0, 0}^{\cj_x\cj_x}(t)$, for the bond currents in the same ($x$) direction as follows:
\begin{align}\label{eq:j0jt_2D_COMII_sum}
    C_{0,0}^{\cj_x\cj_x}(t,0) = \Gamma_{0,0}^{xx}\delta(t) - \frac{\langle m^2 \rangle \mu_1 \mu_2}{64} \frac{1}{L^2} \sum \limits_{q_x, q_y} e^{-D\omega t} \lambda^2_{q_x},
\end{align}
with $\Gamma_{0,0}^{xx}=\mu_2\langle m^2 \rangle/4.$
In the thermodynamic limit and for $t>0$, the above equation can be written in the following integral form
\begin{align}\label{eq:j0jt_2D_COMII_int}
    C_{0,0}^{\cj_x\cj_x}(t,0) &= - \frac{\langle m^2 \rangle \mu_1\mu_2}{64} \frac{1}{(2\pi)^2}\int\limits_0^{2\pi}\int\limits_0^{2\pi}dq_xdq_y e^{-D\omega t}\lambda^2(q_x)\\ \nonumber &\simeq -\frac{\mu_2 \langle m^2 \rangle}{32\pi\mu_1}t^{-3},
\end{align}
where the above asymptotic expression has been calculated for a large time and by using the approximation $\lambda(q) \approx q^2$.

Using Eq. \eqref{eq:j0jt_2D_COMII_sum}, also the power spectrum for instantaneous bond current can be calculated by taking the Fourier transform of the unequal-time bond current correlation function,
\begin{align}
    S_J(f)&= \int\limits_{-\infty}^\infty dt C_0^{\cj_x \cj_x}(t,0)e^{i2\pi ft} \\ \nonumber &= S_J(0) + \frac{\langle m^2\rangle \mu_1 \mu_2}{32DL^2} \sum\limits_{q_x, q_y} \frac{\lambda^2_{q_x}}{\omega}\frac{4\pi^2f^2}{4\pi^2f^2+D^2\omega^2},
\end{align}
where the zero-frequency mode of the power spectrum is given by
\begin{align}
     S_J(0) &= \frac{\mu_2\langle m^2 \rangle}{8}\left[2-\frac{1}{L^2}\sum_{q_x,q_y} \frac{\lambda^2(q_x)}{\omega}\right]\\ \nonumber &=A_1 \equiv \lim_{T\to\infty} \frac{\langle \cq^2(T) \rangle_c}{T}.
\end{align}
It should be noted that the coefficient $A_1$ is actually nonzero for this particular variant in two (and higher) dimensions, signifying that, in temporal domain, time-integrated bond current is not hyperuniform and its variance has a linear growth as a function of time.

\subsection{Correlation between currents along two perpendicular directions.}

\begin{figure*}
    \centering
    \includegraphics[width=0.5\linewidth]{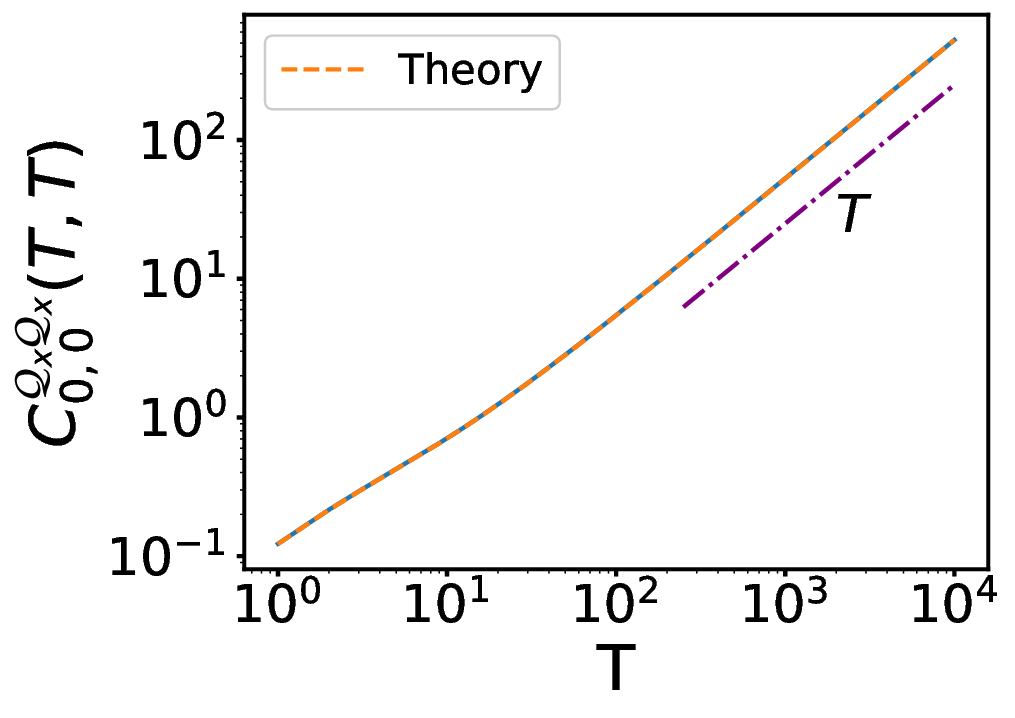}
    \put(-190,140){\textbf{(a)}}
    \includegraphics[width=0.5\linewidth]{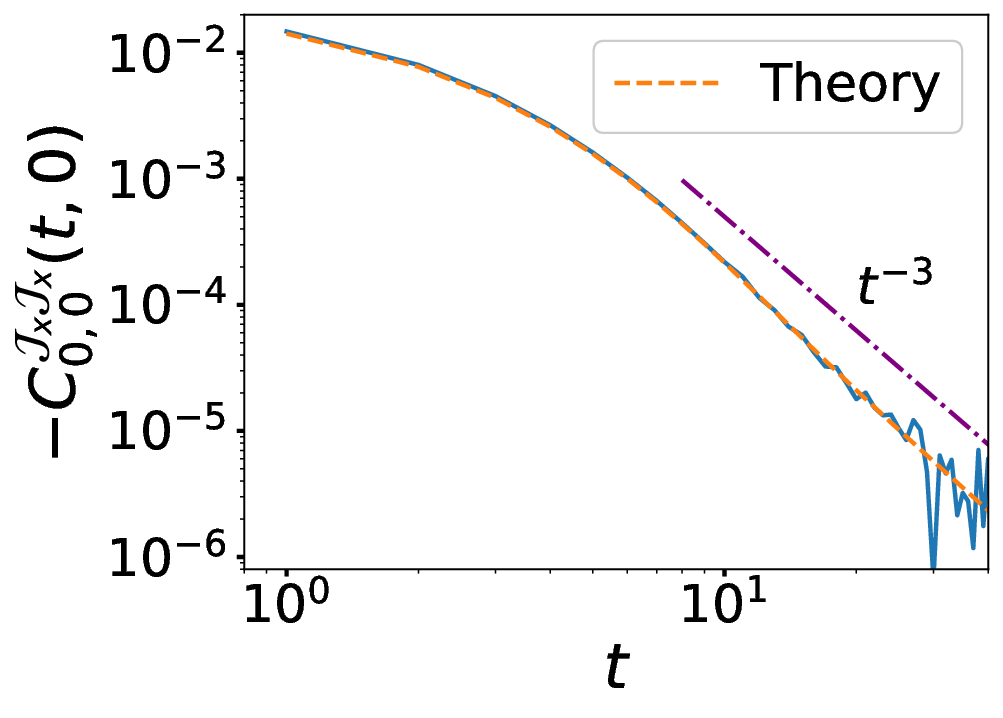}
    \put(-190,140){\textbf{(b)}}\\
    \includegraphics[width=0.5\linewidth]{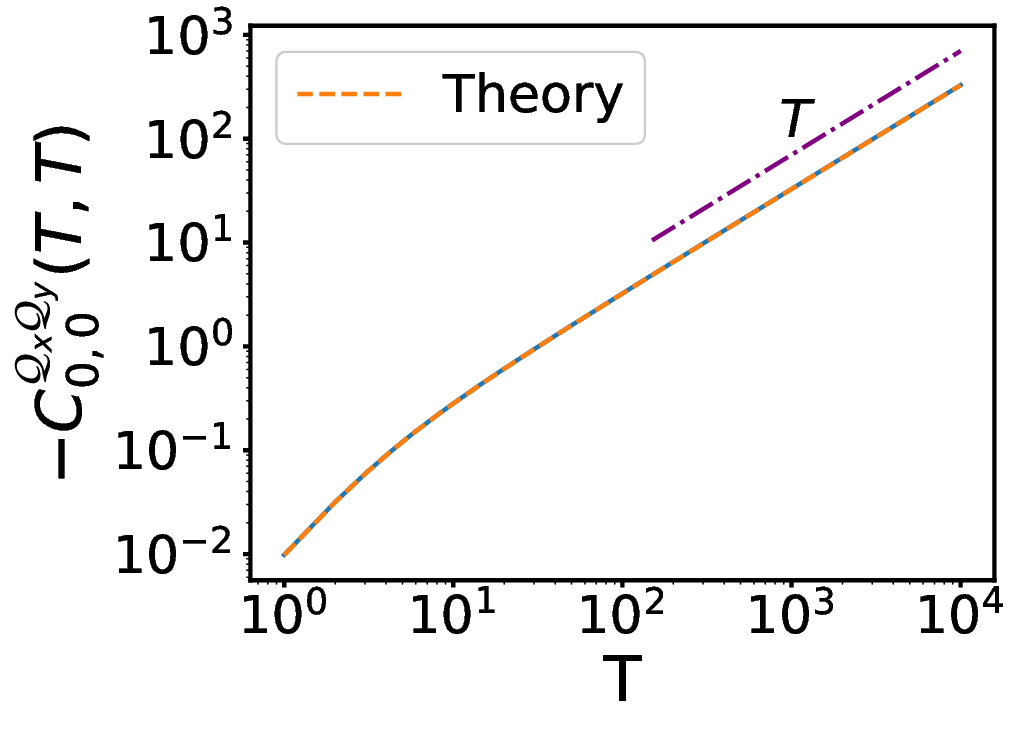}
    \put(-190,140){\textbf{(c)}}
    \includegraphics[width=0.5\linewidth]{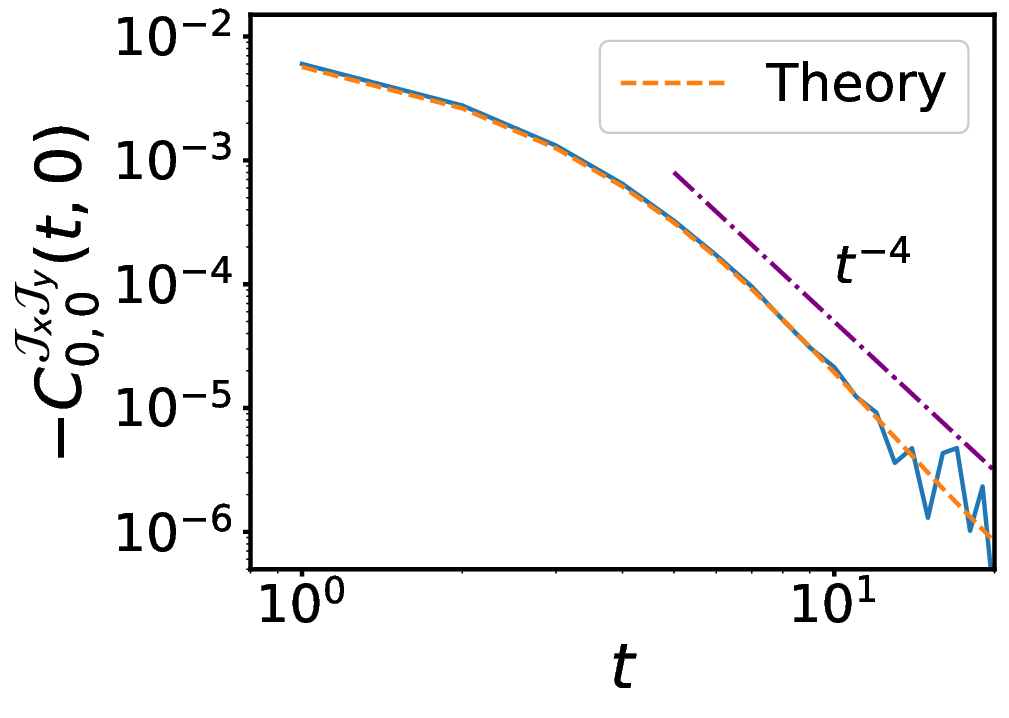}
    \put(-190,140){\textbf{(d)}}
    
    \caption{  {\it MCM-COM II in two dimensions.} Panel (a): The bond-current fluctuation $C_{0,0}^{\cq_x\cq_x}(T,T)$ is plotted as a function of time $T$. Panel (b): Negative of instantaneous current correlation $C_{0,0}^{\cj_x\cj_x}(t,0)$ is plotted as a function of time $t$. Panel (c): The negative of time-integrated cross-current correlation $C_{0,0}^{\cq_x\cq_y}$ is plotted as a function of time $T$.  Panel (d) Negative of instantaneous cross-current correlation $C_{0,0}^{\cj_x\cj_y}(t,0)$ is plotted as a function of time $t$.   The orange dashed lines in panels (a)-(d) represent the theory lines, and the asymptotic (purple dashed-dotted lines) are obtained from Eqs. \eqref{eq:cqxx_2D_COMII}, \eqref{eq:j0jt_2D_COMII_int}, \eqref{eq:cqxyt_2DCOMII_int}, and \eqref{eq:cjxjy_COMII_int}, respectively. We take a periodic system of area $100 \times 100$ and the global density $\rho = 1.0$.}
    \label{fig:MCM_COM_II_2D}
\end{figure*}

Now, we calculate the cross-correlation function for time-integrated bond currents in the two orthogonal directions ($x$ and $y$) as follows:
\begin{align}\label{eq:cqxqy_COMII_sum}
      &C_{0,0}^{\cq_x\cq_y}(T, T)  \\ \nonumber &= -\frac{\langle m^2 \rangle\mu_2}{16}\frac{1}{L^2}\sum_{q_x,q_y}\left [\frac{T}{\omega}-\frac{1-e^{-D\omega T}}{D\omega^2}\right]\Lambda(q_x,q_y)\gamma(q_x,q_y),
\end{align}
where $\gamma(q_x,q_y) = (2\lambda_{q_x} \lambda_{q_y} + \lambda_{q_x}^2-\lambda_{q_y}^2)/(\lambda_{q_x}+\lambda_{q_y})$.
In the thermodynamic limit, we can write the above equation in an integral form,
\begin{widetext}
\begin{align}\label{eq:cqxyt_2DCOMII_int}
   C_{0,0}^{\cq_x\cq_y}(T, T) &= -\frac{\langle m^2 \rangle\mu_2}{16(2\pi)^2}\int\limits_0^{2\pi}\int\limits_0^{2\pi}dq_xdq_y\left [\frac{T}{\omega}-\frac{1-e^{-D\omega T}}{D\omega^2}\right]\Lambda(q_x,q_y)\gamma(q_x,q_y) \\ \nonumber
      & \simeq -\frac{\langle m^2 \rangle\mu_2(10-3\pi)\pi^4}{384}T + \frac{\langle m^2 \rangle\mu_2\pi^2(3\pi-8)}{256\mu_1} -\frac{3\langle m^2 \rangle\mu_2}{128\mu_1^3\pi} T^{-2}, 
      \\ \nonumber
      & \equiv B_1 T + B_2 + B_3 T^{-2},
\end{align}
\end{widetext}
where we have used the approximation $\lambda(q_\alpha) \approx q_\alpha^2$. Note that, in the long-time regime, the cross-correlation of the time-integrated current grows linearly with time, where the prefactor of the growth $B_1 < 0$ is  \textit{negative}. Physically, this negative sign implies that, if some current is generated along $x$ direction over a time interval $T$, the current along $y$-direction flows in the negative direction due to the existence of current loops, or vortices, in two dimensional geometry \cite{Bodineau2008Jun}. However, quite interestingly, the sign of this cross-correlation function $C_{0,0}^{\cq_x\cq_y}$ for orthogonal (along $x$ and $y$) currents is \textit{positive} for the model (two-dimensional MCM-CoMC I) discussed in the previous section.
Similarly, we have calculated the fluctuation of instantaneous current  in a different direction 
\begin{widetext}
\begin{align}\label{eq:cjxjy_COMII_int}
      C_{0,0}^{\cj_x\cj_y}(t, 0) = -\frac{\langle m^2 \rangle\mu_2\mu_1}{128}\frac{1}{(2\pi)^2}\int\limits_0^{2\pi}\int\limits_0^{2\pi}dq_xdq_y e^{-D\omega t}\Lambda(q_x,q_y)\gamma(q_x,q_y)\simeq-\frac{9\langle m^2 \rangle\mu_2}{128\mu_1^3\pi}t^{-4},
\end{align}
\end{widetext}
where we use the approximation $\lambda(q_\alpha)\approx q_\alpha^2$.
In Fig. \ref{fig:MCM_COM_II_2D}, we have plotted various time-dependent quantities involving current correlations in this model. In panel (a) of Fig. \ref{fig:MCM_COM_II_2D}, the time-integrated bond-current fluctuation $C_{0,0}^{\cq_x\cq_x}(T,T)$ is plotted as a function of time $T$. In panel (b) of the same figure, the negative of the instantaneous current correlation $C_{0,0}^{\cj_x\cj_x}(t,0)$ is plotted as a function of time $t$. In panel (c), the negative of time-integrated ``cross"-current correlation $C_{0,0}^{\cq_x\cq_y}$ is plotted as a function of time $T$.  In panel (d), the negative of cross-correlation $C_{0,0}^{\cj_x\cj_y}(t,0)$ for instantaneous bond current is plotted as a function of time $t$. The orange dashed lines in all these panels represent the theory lines, where the asymptotic (purple dashed-dotted lines) are obtained from Eqs. \eqref{eq:cqxx_2D_COMII}, \eqref{eq:j0jt_2D_COMII_int}, \eqref{eq:cqxyt_2DCOMII_int}, and \eqref{eq:cjxjy_COMII_int}, respectively. Here, we have considered a periodic system with area $100 \times 100$ and global density $\rho = 1.0$.

\section{Generalization to arbitrary dimensions}
\label{Sec:dD_MCM_COM}

In the previous sections, we have explicitly calculated the steady-state dynamic correlation functions for instantaneous and time-integrated bond current in MCMs with CoM conservation (MCM-CoMCs) in $d=1$ and $2$ dimensions; moreover, the corresponding exponents governing the decay (or, growth) of these dynamic correlations have been exactly determined. In this section, we discuss how the long-time asymptotics can be generalized to $d$ dimensions through simple dimensional scaling analysis. For example, the (two-point) unequal-time bond-current correlation function $C_0^{\cj_\alpha \cj_\alpha}(t,0)$  has the following asymptotic power-law decay in $d$ dimensions as a function of time $t$, 
\begin{align}\label{eq:cjalpalp}
    C_0^{\cj_\alpha \cj_\alpha}(t,0) \sim -\int\limits_0^\infty q^4e^{-Dq^2t}q^{d-1}dq \sim -t^{-\left(2+{d}/{2} \right)},
\end{align}
where $\alpha \in [x,y]$ denote $x$ or $y$ direction, and we have approximated $\omega({\bf q}) \approx q^2$, with $q$ being the magnitude of the wave-number vector ${\bf q}$ in $d$-dimensions. 
The above asymptotic form has been derived by simple power counting of $q_{\alpha}$'s in the integrands of Eqs. \eqref{eq:j0jt_xx_com_2D_int} and \eqref{eq:j0jt_2D_COMII_int} for the MCM-CoMC models I and II, respectively. 
Alternatively, the expression can also be derived directly from the general expression of current correlations given in Eq. \eqref{eq:cqjjttp_MCM_com_dD} by replacing $\beta$ by $\alpha$ and using the scale transformation $q_{\alpha} = q'_{\alpha}/\sqrt{t}$; for details, see \ref{app:ddim_cal}.  
The power spectrum for the correlation function for the bond currents along a particular direction $\alpha$ in the small-frequency regime has an asymptotic form,
\begin{align}\label{eq:sfalpalph}
    S_J(f) = \int\limits_{-\infty}^\infty dt C_0^{\cj_\alpha \cj_\alpha}(t,0)e^{i2\pi ft} \simeq A_1 +{\rm Const}. f^{\left(1+{d}/{2}\right)},
\end{align}
where $A_1 \equiv S_J(0)$, the zero-frequency mode, can be expressed in the thermodynamic limit as $A_1 = \lim_{T \to \infty}\langle \cq_\alpha^2(T)\rangle / T$. In one dimension, as demonstrated in Secs. \ref{Sec:MCM_CoM_I_1D} and \ref{Sec:1D_MCM_COM_II}, the coefficient $A_1$ vanishes, signifying that the dynamic current fluctuations exhibit ``class-I'' hyperuniformity \cite{Torquato2018Jun}, albeit in the time domain, as previously observed for a different quantity - activity fluctuations - in a sandpile model \cite{Garcia-Millan2018Jul}. However, in higher dimensions ($d > 1$), $A_1$ can be nonzero, depending on the microscopic details (e.g., for MCM CoMC II, $A_1 > 0$, as shown in \eqref{A1-nonzero-2D}). Consequently, dynamic hyperuniformity may be lost even though static fluctuations remain hyperuniform, with $S(q \to 0) \to 0$.

One can also calculate the ``cross''-correlation functions for currents in two orthogonal directions; that is, if a current is initially ($t=0$) generated in a bond in the $x$-direction, one may ask how it is correlated to the current across a bond in (orthogonal) $y$-direction at a later time $t$. The asymptotic form of the cross correlation functions in $d$-dimension can be written as
\begin{align}\label{eq:calpbeta}
    C_0^{\cj_\alpha\cj_\beta}(t,0) \sim -\int\limits_0^\infty q^6e^{-Dq^2t}q^{d-1}dq \sim -t^{-\left(3 + {d}/{2} \right)}.
\end{align}
The above asymptotic form has been derived by power counting of $q_{\alpha}$'s in the integrands of Eqs. \eqref{eq:cjxjy_comI_int} and \eqref{eq:cjxjy_COMII_int} in the MCM-CoMC models I and II, respectively. Alternatively, it can be derived from Eq. \eqref{eq:cqjjttp_MCM_com_dD} by replacing $\beta$ by $\alpha$ and using the transformation $q_{\alpha} = q'_{\alpha}/\sqrt{t}$; for details, see \ref{app:ddim_cal}. 
The asymptotic forms in Eqs. \eqref{eq:cjalpalp}, \eqref{eq:sfalpalph} and \eqref{eq:calpbeta} are expected to be generic for diffusive systems with CoM conservation, regardless of microscopic details; however, the existence of phase transition in the system (e.g., in the presence of a threshold condition in dynamics) can change the scenario. 
As there are no cross-correlations in one dimension, Eq. \eqref{eq:calpbeta} is applicable only for $d>1$.

Equivalently, one can also obtain the variance (second cumulant) $\langle \cq_\alpha^2(T) \rangle_c$ of time-integrated bond-current in a fixed time interval $T$, having the following asymptotic form in the thermodynamic limit,
\begin{align}
   \langle \cq_\alpha^2(T) \rangle_c   \equiv C_{0,0}^{\cq_\alpha \cq_\alpha}(T,T) \simeq A_1 T + A_2 + A_3T^{-{d}/{2}},
\end{align}
where $A_i$'s are density- and parameter-dependent constants. 
The above asymptotic form has been obtained by power counting of $q_{\alpha}$'s in the integrands of Eqs. \eqref{eq:c00tt_com1_2d} and \eqref{eq:cqxx_2D_COMII} in the MCM-CoMC models I and II, respectively. It can also be obtained by taking the inverse Fourier transform of Eq. \eqref{eq:cqQQttp_MCM_com_dD}, replacing $\beta$ by $\alpha$ and then using the scale transformation $q_{\alpha} = q'_{\alpha}/\sqrt{t}$.  
Quite strikingly, depending on dimensions and the microscopic details of the models, the variance of time-integrated bond current saturates at long times, i.e, $A_1=0$ vanishes [see Eqs. \eqref{A1-zero-1D} and \eqref{A1-zero-2D}]. Though this feature is presumably generic in one dimension, however in higher dimensions $d>1$, the coefficient $A_1$ need not always vanish [see Eq. \eqref{A1-nonzero-2D} for MCM-CoMC II in $d=2$]. We can also calculate the cross-correlation function $C_{0,0}^{\cq_\alpha \cq_\beta}(T,T)$, which has an asymptotic form, 
\begin{align}
     C_{0,0}^{\cq_\alpha \cq_\beta}(T,T) \simeq B_1 T + B_2 + B_3T^{-(1+{d}/{2})},
\end{align}
where $B_i$'s are constant (density- and parameter-dependent) coefficients. The asymptotic form has been obtained by power counting of $q_{\alpha}$'s in the integrands of Eqs. \eqref{eq:cqxytt_comI_int} and \eqref{eq:cqxyt_2DCOMII_int} in the MCM-CoMC models I and II, respectively. It can also be derived by taking the inverse Fourier transform of Eq. \eqref{eq:cqQQttp_MCM_com_dD}, using $\alpha \neq \beta$ and then the scale transformation $q_{\alpha} = q'_{\alpha}/\sqrt{t}$. 
As in the case of correlation functions for time-integrated bond currents (in the same direction), the cross-correlation function for currents in the two orthogonal (e.g., $x$ and $y$) directions can also saturate at long times, and $B_1$ in that case is zero (i.e., no long-time linear growth of correlation function), depending on the microscopic details, as demonstrated in the two-dimensional case of MCM-CoMC I [see Eq. \eqref{B1-zero}].
   
\section{Summary and concluding remarks}\label{Sec:Conclusion}

In this paper, we have theoretically investigated steady-state dynamic correlations for currents in a broad class of mass transport processes, known as mass chipping models (MCMs), which have both mass and {\it center of mass} (CoM) being conserved. 
We study these models - MCMs with CoM conservation, called here MCM-CoMCs - specifically in $d=1$ and $2$ dimensional periodic domain of size $L$; the results can be suitably generalized to arbitrary dimensions. 
Notably, the MCM-CoMCs belong to the generalized Kipnis-Marchioro-Pressutti (KMP) models \cite{Kipnis1982Jan, Liu1995Jul, Patriarca2005, Zielen2002Jan, Yakovenko2009Dec, Carinci2013Aug, Das2016Jun}, where masses fragment (i.e, get chipped off from parent masses), diffuse and then coalesce with neighboring masses with constant rate. 
They have broken time-reversal symmetry, thus no detailed balance, in the bulk and, in most cases, their steady states cannot be described by the equilibrium Boltzmann-Gibbs distribution. 
Recently, the role of detailed-balance violation and the CoM conservation have received significant attention in the context of an exotic state of matter, that of ``hyperuniform'' systems, which are examples of many-body correlated systems with their density fluctuations anomalously suppressed. 
Indeed, unlike in typical diffusive systems having a single conserved quantity (e.g., mass), the fluctuation properties of the MCM having both mass and CoM conservation are fundamentally different.
We show that both dynamic and static fluctuations in the MCM-CoMCs are greatly suppressed in spatial and temporal domains, leading to the emergence of an extreme form of  (``class I'') hyperuniform state of matter.

In these models, we show that, in the limit of small wave numbers $q \to 0$, the static structure factor $S(q)$ vanishes as $q^2$. Indeed, this particular dependence of the static structure factor at small wave numbers is a signature of an extreme form of (``class-I" \cite{Torquato2018Jun}) hyperuniformity in the spatial domain, implying an extreme suppression of density fluctuations in the systems. 
Furthermore, we analytically calculate various time-dependent quantities, including the time-varying density profiles that relax from a given initial condition and the dynamic correlation functions for currents.
We show that the two-point dynamic correlation $C_0^{\cj_\alpha \cj_\alpha}(t,0)$ for bond current in a particular direction, with $\alpha \in {x, y}$, varies as $\sim -t^{-\left(2+{d}/{2} \right)}$  as a function of time $t$ in $d$-dimensions. For example, in $d=1$ dimension, the temporal decay is described by a power law $t^{-5/2}$; this should be contrasted with that in the one-dimensional symmetric simple exclusion processes (SSEPs) \cite{Sadhu2016Nov} or the MCMs \cite{Hazra2024Aug}, having only a single (mass) conservation law, where the dynamic correlation function for current decays slower, $\sim t^{-3/2}$, as a function of time. Likewise, in the thermodynamic limit, the power-spectrum $S_J(f)$ for bond current in the models studied here has an asymptotic form as $S_J(f) \simeq A_1 + {\rm Const}.f^{1+d/2}$ for small frequencies $f \to 0$.  That is, in one dimension where we have $A_1=0$, the small-frequency power spectrum for MCM-CoMCs is much suppressed as compared to that in, e.g. the one-dimensional SSEP, where the corresponding behavior is described by the $f^{1/2}$ power law \cite{Sadhu2016Nov}. However, the long-time behavior of the second cumulant, or the variance, $\langle {\cal Q}^2_{\alpha}(T) \rangle_c$ of the time-integrated bond current ${\cal Q}_{\alpha}(T)$ in a time interval $T$ has a long-time asymptotic form $\langle \cq_\alpha^2(T) \rangle_c   \equiv C_{0,0}^{\cq_\alpha \cq_\alpha}(T,T) \simeq A_1 T + A_2 + A_3T^{-{d}/{2}}$, which is quite nuanced and has an intricate behavior, depending on the dimension and microscopic details of the models. While, in one dimension, the variance  $\langle {\cal Q}^2_{\alpha}(T) \rangle_c$, quite strikingly, saturates at long times, the variance in higher dimensions (say, $d=2$) could either saturate or grow linearly with time. Another quantity of interest in higher dimensions is the cross-correlation function $C_0^{\cj_\alpha \cj_\beta}(t,0) \sim -t^{-\left(3+{d}/{2}\right)}$ for currents in two orthogonal (e.g., $x$ and $y$) directions $\alpha$ and $\beta$ (with $\alpha \ne \beta$); the correlation function decays as a power law, which is however faster than that when $\alpha=\beta$. 
The cross-correlation function $C_{0,0}^{\cq_\alpha \cq_\beta}(T,T)$ for currents $\cq_\alpha$ and $\cq_\beta$ in the two orthogonal directions $\alpha$ and $\beta$  (e.g., $x$ and $y$) has a long-time asymptotic form $C_{0,0}^{\cq_\alpha \cq_\beta}(T,T) \simeq B_1 T + B_2 + B_3T^{-(1+{d}/{2})}$, which, depending on the microscopic details, can also saturate at long times, and $B_1$ in that case simply vanishes. The subtle effects due to the formation of current loops (vortices \cite{Bodineau2008Jun}), generated at later times due to the bond current at an initial time ($t=0)$, could physically explain the reduction of current fluctuations, as well as the vanishing of the coefficients $A_1$ and $B_1$ in some cases, in the presence of CoM conservation. 
Usually, in higher dimensions without CoM conservation, these current loops would typically feed in (positive feedback) the current at later times (as well as the cumulative current overall), thus typically enhancing the current fluctuations. However, the CoM conservation generates opposing current loops (clockwise and anticlockwise), which decrease current fluctuations in the systems. Furthermore, depending on the microscopic details, a greater number of opposing current loops can be formed, such as in the case of the first variant MCM-COMC-I, where mass is transported simultaneously along all axes and directions. The precise quantitative characterization of the spatio-temporal structure of these current loops (or, vortices) is, however, beyond the scope of the present study, and will be considered elsewhere \cite{Hazra_unpublished}.

Though MCMs have been extensively investigated throughout the past several decades, the role of an additional conservation law, such as the CoM conservation, on relaxation and fluctuation properties of such systems however remains unexplored and lacks good theoretical understanding. Indeed, recently, there has been growing interest in characterizing large-scale properties of systems having multiple conserved quantities, especially in the context of quantum many-body systems, like fractonic fluids and spin-systems with dipole-like (or, multipole-like) conservation laws \cite{Shenoy2020Feb, Gromov2020Jul}. In the latter cases, the density relaxation is found to be subdiffusive \cite{Morningstar2020Jun, Han2024Mar};  however, it must be noted that these models, of course, have time-reversal symmetry, lack of which, as we have demonstrated here, can indeed have a drastic impact on the relaxation processes. To systematically investigate the issue, here we ask what happens when the time-reversal symmetry, or detailed balance, is actually broken in the bulk, which is the case for the broad class of models studied in this work. Indeed, in the context of the MCM-CoMCs considered here, we find that, despite highly constrained microscopic dynamics with CoM conservation, the density relaxation in the absence of time-reversal symmetry need not be subdiffusive \cite{Bertrand2019Dec, Guo2022Oct, Mukherjee2024Aug}, and we show that the density relaxation in MCM-CoMCs is diffusive. Nevertheless, the fluctuation properties of MCM-CoMCs differ significantly from those of typical diffusive systems with a single conserved quantity, such as symmetric exclusion processes \cite{Arita2014Nov, Sadhu2016Nov} and MCMs without CoM conservation \cite{Hazra2024Aug}.

We conclude the paper with a few remarks.
Despite having a nonequilibrium steady state, which is not described by the Boltzmann-Gibbs distribution and has a nontrivial steady-state structure having finite spatial correlations in the bulk, MCM-CoMCs are amenable to analytical calculations. The analytical tractability arises from the fact that not only the MCM-CoMCs have a ``gradient property'' \cite{Arita2014Nov}, but also the bulk-diffusion coefficient is {\it independent} of density; in other words, the local diffusive current is a {\it linear} function of local masses as seen in Eq. \eqref{JD}. In fact, the latter aspect - the {\it linearity property} of Eq. \eqref{JD} - is quite crucial for performing explicit analytical calculations as done in the present work. We show that the above linearity property leads to the time-evolution equations for the two-point (in general, $n$-point) correlation functions to {\it not} involve higher-order correlations and thus close onto themselves. Of course, many other dynamical variations of these models are possible. For example, adding a threshold condition on the mass transfer rules (in other words, mass-dependent chipping rates) would certainly make the model more interesting because of the possibility of a nonequilibrium phase transition in the system, similar to the absorbing phase transition observed in the BRO model \cite{Corte2008May} or the continuous-mass version of the celebrated Manna sandpile \cite{Manna1991Apr, Dickman2001Oct}. However, it would immediately make the correlation functions have an infinite BBGKY-like hierarchy, rendering the model analytically almost intractable in even one dimension. Indeed, it remains a challenge to derive rigorous results for time-dependent properties of the latter class of systems, and the exact results derived in this work could provide some useful insights in that direction.

\section*{Acknowledgement}

We thank  Salvatore Torquato and R. Rajesh (during a conference ``STATPHYS Kolkata XII'', 2023) for their comments, which motivated this study. We thank Deepak Dhar for careful reading of the manuscript and useful discussions.


\appendix
\section{Dynamic correlations in $d$ dimensions: Sketch of the calculations} \label{app:ddim_cal}
Let us consider a system on a $d$-dimensional periodic square lattice, where masses diffuse with both mass and CoM conserved. Now let us denote the time-integrated bond currents as $\cq_\alpha$ and $\cq_\beta$ along two orthogonal directions $\alpha$ and $\beta$ (like $x$ and $y$ for $d=2$), respectively. In this case, the unequal-time ($t>t'$) current correlation function in Fourier space can be written as 
\begin{align}
\label{eq:cqqttp_d}
    \frac{d}{dt}\tilde{C}^{\cq_\alpha\cq_\beta}_{\textbf{q}}(t,t')=D(1-e^{iq_\alpha})\tilde{C}^{m\cq_\beta}_{\textbf{q}}(t,t'),
\end{align}
where $\textbf{q}$ is $d$-dimensional  wave-number vector with $q_\alpha$ being the component of $\textbf{q}$ along direction $\alpha$.
Equation \eqref{eq:cqqttp_d} has a similar structure as that in one dimension, with $D$ being the bulk-diffusion coefficient.
To solve the above equation, we require the unequal-time mass-current correlation, which satisfies the following equation:
\begin{align}\label{eq:cmqttp_d}
    \frac{d}{dt}\tilde{C}^{m\cq_\beta}_{\textbf{q}}(t,t')=-D\omega(\textbf{q})\tilde{C}^{m\cq_\beta}_{\textbf{q}}(t,t'),
\end{align}
where $\omega(\textbf{q})=\sum_\alpha \lambda(q_\alpha)$. 
To obtain the general solution of the above two equations, namely Eqs. \eqref{eq:cqqttp_d} and \eqref{eq:cmqttp_d}, we also require the equal-time current-current and mass-current correlation functions. First, we write the time evolution of equal-time current-current correlation function as
\begin{align}\label{eq:qqtt_d}
    \frac{d}{dt}\tilde{C}^{\cq_\alpha\cq_\beta}_{\textbf{q}}(t,t)&=D(1-e^{iq_\alpha})\tilde{C}^{m\cq_\beta}_{\textbf{q}}(t,t)\\ \nonumber &+D(1-e^{iq_\beta})\tilde{C}^{m\cq_\alpha}_{\textbf{q}}(t,t)+\tilde{\Gamma}^{\alpha\beta}_{\textbf{q}}(\rho),
\end{align}
where $\tilde{\Gamma}^{\alpha\beta}_{\textbf{q}}(\rho)$ is a model-dependent quantity. The equal-time mass-current correlation can be expressed as
\begin{align}\label{eq:mqtt_d}
     \frac{d}{dt}\tilde{C}^{m\cq_\beta}_{\textbf{q}}(t,t)=-D\omega(\textbf{q})\tilde{C}^{m\cq_\beta}_{\textbf{q}}(t,t)+\tilde{f}^\beta_{\textbf{q}}(\rho),
\end{align}
where 
\begin{align}
    \tilde{f}^\beta_{\textbf{q}}(\rho) = (1-e^{-iq_\beta})G^{I/II}_\textbf{q}(\rho).
\end{align}
Here $G^{I}_\textbf{q}(\rho) = [D\tilde{C}^{mm}_\textbf{q} (\rho) - g^{I}(\rho)\sum_{\alpha} \lambda(q_{\alpha})]$  and $G^{II}_\textbf{q}(\rho) = [D\tilde{C}^{mm}_\textbf{q}(\rho)  - g^{II}(\rho) \sum_{\alpha} \lambda(q_{\alpha}) \delta_{\alpha,\beta}]$, and $g^{I}(\rho)$ and $g^{II}(\rho)$ are two model-dependent constants; also, in the above equation, case $I$ corresponds to simultaneous mass transfer along all $d$ axes and case $II$ corresponds to mass transfer along a $\beta$-axis (chosen randomly during a mass transfer event), whereas CoM is conserved in both cases.
By integrating Eq. \eqref{eq:mqtt_d}, we obtain
\begin{align}\label{eq:mqtptp_d}
     \tilde{C}^{m\cq_\beta}_{\textbf{q}}(t',t')=\int\limits_{0}^{t'}dt''e^{-D\omega_{\textbf{q}}(t'-t'')}\tilde{f}^\beta_{\textbf{q}}(\rho).
\end{align}
Then by integrating Eq. \eqref{eq:qqtt_d} and  using Eq. \eqref{eq:mqtptp_d}, finally we obtain the equal-time current correlation function in the Fourier space, 
\begin{widetext}
\begin{align}\label{eq:cqabtt_d}
     \tilde{C}^{\cq_\alpha\cq_\beta}_{\textbf{q}}(t,t) =\int\limits_{0}^{t}dt\tilde{\Gamma}^{\alpha\beta}_{\textbf{q}}(\rho)+D\int\limits_{0}^{t}dt'\int\limits_{0}^{t'}dt''[\lambda(q_\alpha)+\lambda(q_\beta)-\lambda(q_\alpha-q_\beta)]G^{I/II}_\textbf{q}(\rho)e^{-D\omega_\textbf{q}(t'-t'')}.
\end{align}
Now integrating Eq. \eqref{eq:cqqttp_d} and using Eqs. \eqref{eq:cqabtt_d}, \eqref{eq:cmqttp_d} and \eqref{eq:mqtptp_d}, finally we obtain the unequal-time current correlation in the Fourier space as
\begin{align}\label{eq:cqQQttp_MCM_com_dD}
    \tilde{C}_\textbf{q}^{\cq_\alpha\cq_\beta}(t,t^\prime)= \tilde{\Gamma}_\textbf{q}^{\alpha\beta} t^\prime +D\Lambda(q_\alpha,q_\beta)G^{I/II}_\textbf{q}(\rho)\Bigg[\int\limits_0^{t^\prime}dt^{\prime\prime}\int\limits_{0}^{t^{\prime\prime}}dt^{\prime\prime\prime}e^{-D\omega_\textbf{q}(t^{\prime\prime}-t^{\prime\prime\prime})} +\frac{1}{2}\int\limits^t_{t^\prime}dt^{\prime\prime}\int\limits_{0}^{t^{\prime\prime}}dt^{\prime\prime\prime}e^{-D\omega_\textbf{q}(t^{\prime\prime}-t^{\prime\prime\prime})} \Bigg],
\end{align}
\end{widetext}
where $\Lambda(q_\alpha,q_\beta)=[\lambda(q_\alpha)+\lambda(q_\beta)-\lambda(q_\alpha-q_\beta)]$. 
For $t>0$, the unequal-time (instantaneous) bond current correlation function can be written from the above equation as 
\begin{align}\label{eq:cqjjttp_MCM_com_dD}
    &C_{\textbf{r}=0}^{\cj_\alpha\cj_\beta}(t, 0) =  \frac{D}{2}\int_\textbf{q} d^{d}\textbf{q} \Lambda(q_\alpha,q_\beta) G^{I/II}_\textbf{q} e^{-D\omega_\textbf{q}t}.
\end{align}
Furthermore, Eqs. \eqref{eq:cqQQttp_MCM_com_dD} and \eqref{eq:cqjjttp_MCM_com_dD} can also be applied to obtain the current correlations along the same direction, say $\alpha$, by replacing $\beta$ with $\alpha$ in the factor $\Lambda$.
Now, we have, in the small wave-number limit,
\begin{align}
\Lambda(q_\alpha, q_\beta) \simeq
    \begin{cases} 
        \frac{q_\alpha^2 q_\beta^2}{2} & \text{for } \alpha \neq \beta, \\
        2q_\alpha^2 & \text{for } \alpha = \beta 
    \end{cases}
\end{align}
and $G^{I/II}_\textbf{q} \sim {\cal O}(q_\alpha^2, q_{\beta}^2)$; these can be used into \eqref{eq:cqjjttp_MCM_com_dD} to derive the asymptotic expression for the bond current correlation function as given in the main text in Eqs. \eqref{eq:cjalpalp} and \eqref{eq:calpbeta}.


\bibliography{ref.bib}


\end{document}